\documentclass[twocolumn,twocolappendix]{aastex63}

\usepackage{graphicx}
\usepackage{amsmath}
\usepackage{amssymb}

\usepackage{tabularx}
\usepackage{natbib}

\usepackage[maxfloats=256]{morefloats}
\DeclareUnicodeCharacter{2212}{-}
\usepackage{enumerate}

\newcommand{\Mstar}{\mbox{$\mathrm{M_{\ast}}$}}
\newcommand{\Mdyn}{\mbox{$\mathrm{M_{Dyn}}$}}

\newcommand{\Morph}{\mbox{$\mathrm{Morph.}$}}
\newcommand{\SFRssp}{\mbox{$\mathrm{SFR_{ssp}}$}}
\newcommand{\Reff}{\mbox{$\mathrm{R_{eff}}$}}
\newcommand{\Mmol}{\mbox{$\mathrm{M_{H2}}$}}
\newcommand{\MmolInf}{\mbox{$\mathrm{M_{H2}^{inf}}$}}
\newcommand{\MmolExt}{\mbox{$\mathrm{M_{H2}^{ext}}$}}

\newcommand{\MHI}{\mbox{$\mathrm{M_{HI}}$}}
\newcommand{\hi}{\mbox{\rm H{\small I}}}
\newcommand{\FmolInf}{\mbox{$\mathrm{F_{H2}^{inf}}$}}
\newcommand{\FmolExt}{\mbox{$\mathrm{F_{H2}^{ext}}$}}
\newcommand{\Fmol}{\mbox{$\mathrm{F_{H2}}$}}

\newcommand{\Fgastot}{\mbox{$\mathrm{F_{gas}}$}}
\newcommand{\FgastotI}{\mbox{$\mathrm{F_{gas}^{inf}}$}}
\newcommand{\FgastotE}{\mbox{$\mathrm{F_{gas}^{ext}}$}}

\newcommand{\Cindex}{\mbox{$\mathrm{C}$}}
\newcommand{\ba}{\mbox{$\mathrm{b/a}$}}
\newcommand{\DL}{\mbox{$\mathrm{D_L}$}}
\newcommand{\ioniz}{\mbox{$\mathrm{U}$}}
\newcommand{\redshiftz}{\mbox{$\mathrm{z}$}}

\newcommand{\incli}{\mbox{$\mathit{i}$}}
\newcommand{\SFMS}{\mbox{${\Delta{\mathrm{SFMS}}}$}}
\newcommand{\Tfifty}{\mbox{$\mathrm{T50}$}}
\newcommand{\Teighty}{\mbox{$\mathrm{T80}$}}
\newcommand{\Tninety}{\mbox{$\mathrm{T90}$}}
\newcommand{\Tefty}{\mbox{$\mathrm{{T80}/{T50}}$}}
\newcommand{\Tnfty}{\mbox{$\mathrm{{T90}/{T50}}$}}

\newcommand{\EWre}{\mbox{$\mathrm{EW(H\alpha)}$}}
\newcommand{\DeltMZR}{\mbox{$\Delta{\mathrm{MZR}}$}}
\newcommand{\SFE}{\mbox{$\mathrm{SFE}$}}
\newcommand{\SFEExt}{\mbox{$\mathrm{SFE}^{ext}$}}
\newcommand{\SFEInf}{\mbox{$\mathrm{SFE}^{inf}$}}
\newcommand{\sSFR}{\mbox{$\mathrm{sSFR}$}}
\newcommand{\lambdaRe}{\mbox{$\mathrm{\lambda}$}}

\newcommand{\ha}{\ensuremath{\rm H\alpha}}
\newcommand{\hb}{\ensuremath{\rm H\beta}}
\newcommand{\othree}{\textrm{[O\,{\sc iii}]}}
\newcommand{\otwo}{\textrm{[O\,{\sc ii}]}}

\newcommand{\ntwo}{\textrm{[N\,{\sc ii}]}}

\newcommand{\stwo}{\textrm{[S\,{\sc ii}]}}

\accepted{}
\submitjournal{ApJ}

\shorttitle{Which galaxy property is the best gauge of the oxygen abundance?}
\shortauthors{Alvarez-Hurtado et al.}

\graphicspath{{./}{figures/}}
\begin{document}


\title{Which galaxy property is the best gauge of the oxygen abundance?}

\correspondingauthor{Alvarez-Hurtado, P.}
\email{palvarez@astro.unam.mx}

\author{Alvarez-Hurtado, P.}
\affiliation{Instituto de Astronom\'ia, Universidad Nacional Aut\'onoma de M\'exico, A.P. 70-264, 04510 M\'exico, CDMX, M\'exico}

\author{Barrera-Ballesteros,J.K.}
\affiliation{Instituto de Astronom\'ia, Universidad Nacional Aut\'onoma de M\'exico, A.P. 70-264, 04510 M\'exico, CDMX, M\'exico}

\author{S\'anchez, S.F.}
\affiliation{Instituto de Astronom\'ia, Universidad Nacional Aut\'onoma de M\'exico, A.P. 70-264, 04510 M\'exico, CDMX, M\'exico}

\author{Colombo, D.}
\affiliation{Max-Planck-Institut für Radioastronomie, Auf dem Hügel 69, 53121 Bonn, Germany} 

\author{L\'opez-S\'anchez, A.R.}
\affiliation{Australian Astronomical Observatory, 105 Delhi Road, North Ryde, NSW 2113, Australia}

\affiliation{Department of Physics and Astronomy, Macquarie University, NSW 2109, Australia}

\affiliation{Australian Research Council Centre of Excellence for All Sky Astrophysics in 3 Dimensions (ASTRO 3D), Australia}

\author{Aquino-Ort{\'\i}z,E.}
\affiliation{Instituto de Astrof\'isica, Pontificia Universidad Cat\'olica de Chile, Av. Vicuña Mackenna 4860, 782-0436 Macul, Santiago, Chile}
%


\begin{abstract}

We present an extensive exploration of the impact of 29 physical parameters in the oxygen abundance for a sample of 299 star-forming galaxies extracted from the extended CALIFA sample. We corroborate that the stellar mass is the physical parameter that better traces the observed oxygen abundance (i.e., the mass-metallicity relation, MZR), while other physical parameters could play a potential role in shaping this abundance, but with a lower significant impact. We find that the functional form that best describes the MZR is a {third-order polynomial function}. From the residuals between this best functional form and the MZR, we find that once considered the impact of the mass in the oxygen abundance, the other physical parameters do not play a significant secondary role in shaping the oxygen abundance in these galaxies (including the gas fraction or the star formation rate). Our analysis suggests that the origin of the MZR is related to the chemical enrichment evolution of the interstellar medium due, most likely, to the build-up of stellar mass in these star-forming galaxies. 

\end{abstract}


\section{Introduction}
\label{sec:intro}

One of the most significant problems in extragalactic astrophysics { is understanding the physical processes which determine the observed
chemical distribution} in the nearby Universe. One of the most common ways to estimate the oxygen abundance in galaxies is through its empirical or theoretical calibrators using strong optical emission lines from the ionized gas in the interstellar medium (ISM). 

The flux from these emission lines reflects the chemical composition of the ISM, which is a consequence of the enrichment by the stellar population, modulated by the inflow and outflow of gas \citep[e.g.,][]{Garnett02,Tremonti04,Finlator08,Spitoni10}. { Thus, a relation is expected between} those properties and the chemical composition of the ISM. \citet{mcclure68,Lequeux79} { were among the first to point out} an increase of the oxygen abundance { with the absolute magnitude} in a sample of irregular and compact galaxies. Later studies found similar trends between the oxygen abundance and different parameters of galaxies, like their morphology \citep[e.g.,][]{1981ApJ...243..127K, 1984AJ.....89.1300B, Skillman89,1992MNRAS.259..121V,Zaritsky94,Calura09}, and other galaxy properties \citep[e.g., molecular and atomic gas fractions, and kinematic properties ][]{1992MNRAS.259..121V,Zaritsky94,Garnett02}. Considering the absolute magnitude as a tracer of the stellar mass, the relation derived by \cite{Lequeux79} implies the existence of a relation between the oxygen abundance and the stellar mass. Indeed, \citet{Tremonti04}, hereafter T04, derived the first systematic exploration of this relation, known as the Mass-Metallicity relation (MZR)\footnote{Along this article, we refer to oxygen abundance ($\mathrm{12+log(O/H)}$) as gas phase metallicity or simply metallicity}, for $\sim$53,400 star-forming galaxies at $0.005 < \mathrm{z} < 0.25$ included in the the Sloan Digital Sky Survey  \citep[SDSS,][]{2000AJ....120.1579Y}. They showed that these two parameters follow a tight relation ($\sigma_{log(O/H)}\sim$0.1 dex), in which the average oxygen abundance increase with stellar mass, reaching a plateau above  $\sim\,10^{10.5} \mathrm{M_{\odot}}$. 
The above studies { have posited that} the parameter that best describes the observed oxygen abundance in the ISM is the stellar mass. However, there is scarce literature that explores the correlation between a comprehensive set of { galaxy properties} and the oxygen abundance.

The shape of the MZR has been described using multiple functional forms, including  a linear -- using the absolute magnitude \citep[e.g., ][]{1984ApJ...281L..21R,Berg12} or with the stellar mass \citep[e.g.,][]{2020A&A...634A.107Y}, exponential \citep[e.g.,][]{Kewley2008,Sanchez17,Sanchez19}, or even high-order polynomials \citep[e.g.,][]{Kewley2008,Mannucci10,RosalesOrtega12}.
We note that these functional forms have been adopted in order to provide the best representation of the observed MZRs. Thus, it is expected that these forms vary depending on the data, and even on the adopted abundance calibrator \citep[e.g., ][]{Kewley2008,Curti_2017,Curti20}. However, these studies used single aperture spectroscopic data, which limits the information to an integrated (average) quantity within a given aperture (its physical size varies with the distance of each galaxy). This can be circumvented by the use of integral-field spectroscopic data over large galaxy samples, as illustrated in \citet{S2020}.

Besides the study of the shape of the MZR, different authors have explored the possible secondary relations with other galaxy parameters. For instance, { \citet{Mannucci10} and \citet{LaraLopez2010}, proposed a} secondary relation with the star-formation in which galaxies with high star-formation rate (SFR) at a fixed stellar mass present lower metallicities. More recently, other authors have claimed the existence of a secondary relation with the both the gas fraction \citep[e.g.,][]{Brooks_2007,2016MNRAS.456.2140M,2019MNRAS.484.5587T}, and the atomic or molecular gas content \citep[e.g.,][]{Hughes13,2013MNRAS.433.1425B, 2016A&A...595A..48B, 2021ApJ...908..226H} in galaxies. However, different authors have proposed that this secondary relation may depend on the calibrator \citep[e.g.,][]{Kewley2008,2016ApJ...823L..24K,Telford16}, on the aperture bias of the SDSS spectroscopic data \citep[e.g.,][]{Sanchez13, BB18}, { and it may} not be as significant or fundamental as initially proposed.
{ The advent of angular-resolved spectroscopic observations (or Integral Field Spectroscopy, IFS) in large samples of galaxies have allowed the { exploration of} the MZR at a given characteristic size of the galaxies (e.g., effective radius). These studies have been performed on the IFS surveys with the largest samples such as: the Calar Alto Legacy Integral Field Area Survey \citep[CALIFA,][]{Sanchez13}, the Mapping Galaxies at Apache Point Observatory Survey \citep[MaNGA,][]{Bundy_2014}, and the Sydney-AAO Multi-object Integral field spectrograph survey \citep[SAMI,][]{Croom_2012}. As main result, these studies have questioned the existence of such secondary relationship with SFR regardless of the abundance calibrators and the aperture size of the observation technique \citep[e.g.,][]{BB17,Sanchez17,Sanchez19}.} Moreover, several studies attempt to explore the shape of the MZR using different approaches \citep[e.g.,][]{Ellison08,Sanchez19}. For instance, \citet{Sanchez19} showed that the use of different functional forms may produce very different residuals that may lead to the apparent presence of secondary relations.

Therefore, the characterization of the MZR in the local universe is fundamental if we want to understand the physical processes that drive this relation, providing a possible anchor point when exploring whether some other physical parameter drives the observed oxygen abundance.

To overcome the aforementioned issues, we present in this study a comprenhensive exploration of the correlation, or lack of thereof, between a large set of physical galaxy's parameters and the oxygen abundance measured at a given characteristic distance (its effective radius) using a heterogeneous set of calibrators thanks to the IFS dataset provided by the CALIFA survey \citep{Sanchez12}. Furthermore, we explore with this dataset the functional form, and different statistical treatments, that best describes the observed MZR for our dataset. Using the residuals of the MZR with respect to the best functional form, we further explore whether those galaxy's parameters still have a significant impact in the MZR.

The paper is structured as follows. In Sec.~\ref{sec:data} we present an overview of our dataset. In Sec.~\ref{ref:Results} we also explore in detail the correlations between different galaxy's parameters and the oxygen abundance as well as the functional form that best describes the MZR; we also explore the impact on the residuals of the MZR from those parameters. We discuss our main results in Sec.~\ref{sec:Discussion}, with a summary of the main conclusions in Sec.~\ref{sec:Conclusions}. Finally, we present a brief description of the calibrators { employed in} this article in App.~\ref{app_calib}. Throughout this manuscript, we assume a Hubble parameter of $H_0=71 (km/s)/Mpc$ and $\Lambda$CDM concordance cosmology with $\Omega_m=0.3, \Omega_{\Lambda}=0.7$.

\section{Sample and Data}
\label{sec:data}

\subsection{The eCALIFA survey} 
\label{sec:eCALIFA_data}

{ The eCALIFA survey is} an extension of the CALIFA survey \citep[extended Calar Alto Legacy Integral Field Area,][]{Sanchez12,Lacerda20}, that increases the original $\sim$600 observed galaxies up to $\sim$1000. This extended sample covers { a similar footprint to} the original CALIFA mother sample. However, some of the selection criteria were slightly relaxed to increase galaxy types  with low numbers in the original sample \citep[e.g., dwarfs, large ellipticals, and SN hosts][]{Sanchez16,Lacerda20}. In particular, the main sample selection, a diameter cut that ensures that galaxies are sampled up to 2.5\Reff \citep{walcher14} was fully considered, conserving the same original exposure time and instrumental setup \citep{EspinosaP20}.

These observations were performed using the Potsdam Multi-Aperture Spectrophotometer \citep[PMAS][]{Roth} employing the PPAK configuration \citep{kelz06} at the Calar Alto 3.5m telescope. This { setup provides a science} field-of-view (FoV) of 74\arcsec $\times$ 64\arcsec, that is sampled with a 100\% covering factor and a 2.5\arcsec$/$FWHM resolution by adopting { a 3-pointing dithering scheme} \citep{Sanchez12,pipe3d_ii}.  { For this study, we} use the low spectral resolution configuration adopted by the CALIFA project (V500) which includes wavelengths from $\sim$ 3745 \AA\, to 7500 \AA, covering a wide range of the { most relevant strong emission lines} for studying the ionized gas (from \otwo $\lambda$ 3727 to \stwo $\lambda $ 6731). The details of the CALIFA project, the sample and data reduction are explained in detail in \citet{Sanchez12}. The reduced dataset for each galaxy consists of a datacube in which the $x$ and $y$ axis represent the projected distribution of the galaxy in RA and DEC, and the $z$-axis corresponds to the wavelength coverage. 

\begin{figure}
\centering
	\includegraphics[width=\columnwidth]{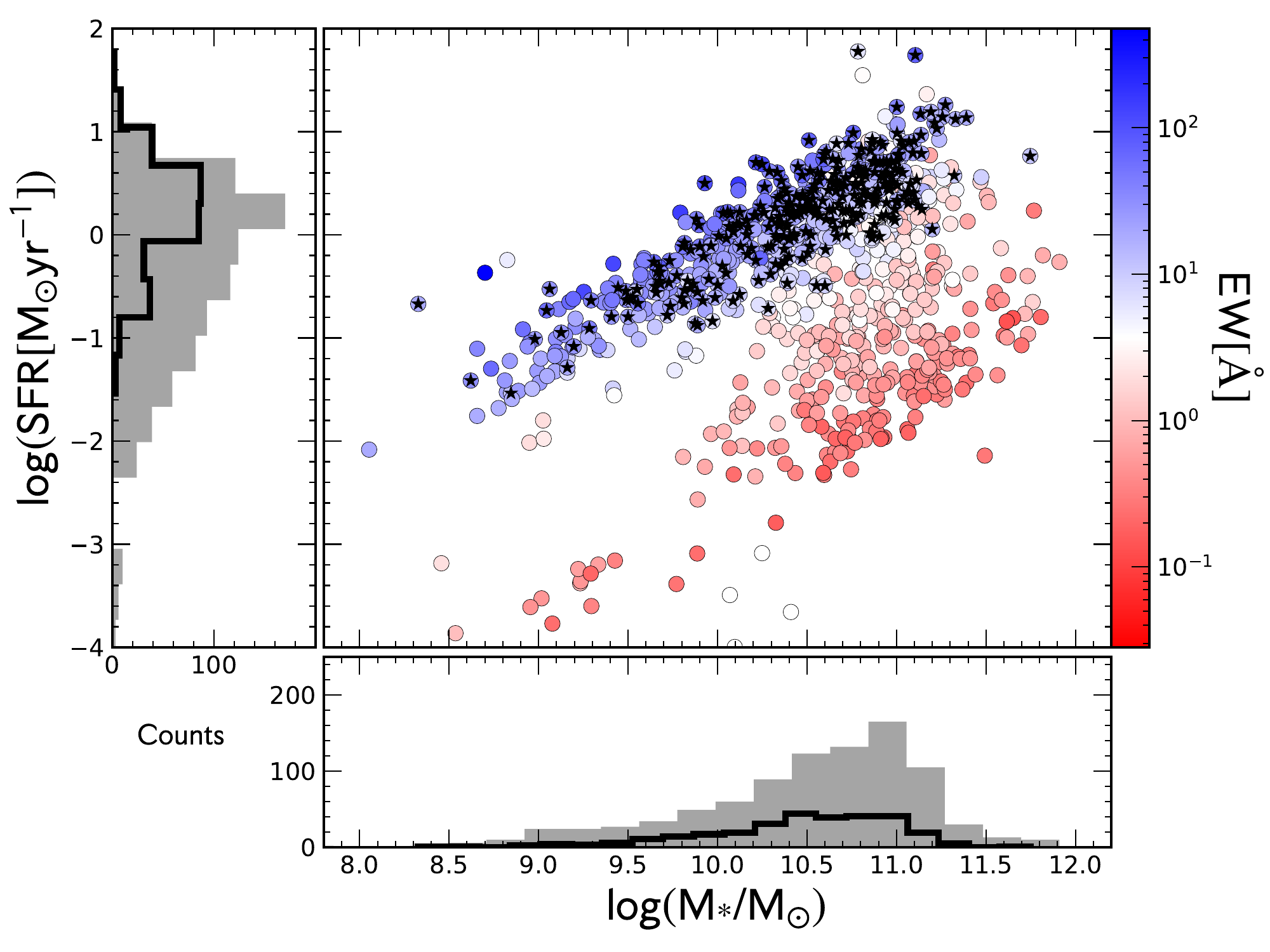}
	\caption{SFR against \Mstar\ for the eCALIFA survey color-coded according to the EW(\ha). Each color-coded circle represents a galaxy in the eCALIFA sample. The black stars correspond to the sub-sample of star-forming galaxies selected to explore the MZR. The adjacent panels in the $y$ and $x$ axis show the distribution of the SFR and \Mstar, respectively. The grey-filled histograms correspond to the complete sample of the eCALIFA galaxies. The sample of star-forming galaxies selected to analyze the MZR are shown as a black histogram.
	}
\label{fig:logM_Morph}
\end{figure}

The science datacubes are then analyzed using the \textsc{Pipe3D} pipeline \citep{pipe3d,pipe3d_ii}. This tool { provides a set} of 2D maps comprising the spatial distribution { of a set} of physical parameters. For instance, { for given emission line,} the pipeline provides a map of its integrated flux, systemic velocity, velocity dispersion, and equivalent width. Similar maps { are derived
} for the properties of the stellar population by deriving the best fit from a library of templates of single stellar populations (SSP). For each parameter, this pipeline also generates a map with its corresponding error estimation \citep[see a detailed description of the pipeline in ][]{pipe3d,pipe3d_ii}. This pipeline has been extensively used in the analysis of different IFS datasets, including but not limited to CALIFA \citep[e.g.,][]{Sanchez16,2019MNRAS.482.1427G,LopezCoba19,EspinosaP20,2020MNRAS.499.4838M,Camps21}, SAMI \citep[e.g.,][]{Sanchez19}, MaNGA \citep[e.g.,][]{2016MNRAS.463.2799I,BB17,2020A&A...644A.117L,2021MNRAS.503.1345S} and MUSE \citep[e.g.,][]{LopezCoba20}, and tested using mock IFS data created from hydro-dynamical simulations \citep[e.g.,][]{Ibarra-Medel19}. { In Fig.~\ref{fig:logM_Morph} we represent the eCALIFA sample in the SFR-\Mstar plane color coded by their EW(\ha). The eCALIFA sample provides a significant coverage of this plane.}

\subsection{Estimation of oxygen abundance} 
\label{sec:oxy_abu}

We are primarily interested in the estimation of the characteristic oxygen abundance for each galaxy. For doing so, we follow \citet{Sanchez14} and \citet{Sanchez19}. They adopt as the characteristic abundance of a galaxy the value 
estimated at the effective radius.

We rely on strong-line calibrators that make use of the following emission lines \ha, \hb, \otwo $\lambda$3727, \othree $\lambda$4959, \othree $\lambda$5007, \ntwo $\lambda$6548, and \ntwo $\lambda$6583. Based on the two-dimensional distributions { provided for those}  emission lines by Pipe3D, we derive the corresponding oxygen abundance { maps, after selecting those } regions which ionization is compatible with young massive OB stars (i.e., related to a recent star-formation event). This selection is based on the prescriptions outlined in \cite{Sanchez21}. Thus, we use only those regions in which the \othree/\hb\ and \ntwo/\ha\ line ratios are below the Kewley demarcation curve { \citep{Kewley01}} with an EW(\ha)$>$6\AA. {  We derive the oxygen abundance for the selected regions using} a set of calibrators (see Appendix \ref{app_calib}, for the full list). { We adopt O3N2 (as calibrated by \cite{marino13}) as our fiducial calibrator (hereafter designated as M13-O3N2).}. Once we derive the oxygen abundance two dimensional map for each galaxy, we derive its radial distribution by performing an azimuthal average following the position angle and ellipticity of each galaxy. Then, we fit a single linear relation to this radial distribution within 0.5-2.0 \Reff, to derive the oxygen abundance at the effective radius \citep[following][]{2006ApJ...651..155M,Sanchez13,2014A&A...562A..47G,Sanchez16}. We address the possible metallicity calibration discrepancy by incorporating a similar analysis using a set of six different calibrators \citep[we refer them as { D16-N2S2}, { PP04-O3N2}, { D13-PYQZ}, { P12-T2} and { M13-N2}, respectively][]{dop16,pp04,PYQZ,T2,marino13}. Each calibrator is created based on different assumptions (see details in Appendix \ref{app_calib}). Hence, each calibrator presents different uncertainties (e.g., { M13-O3N2} $\sim0.08$ dex). Moreover, each calibrator has a different dependence on different emission lines, and therefore, it may present different dependencies on physical properties, including the dust attenuation, the ionizing source, and the ionization parameter, among others. For instance, if any possible secondary relation is observed only on a particular set of calibrators, it may depend on the calibrator's physical assumptions (i.e., an induced relation by how the calibrator was derived). When necessary we describe the differences in the results depending on the adopted calibrator. 

\subsection{Estimation of integrated galaxy's properties} 
\label{sec:galaxy_prop}

{
For each galaxy, we derive the total stellar mass, \Mstar, by integrating the stellar mass surface density ($\Sigma_{\ast}$) spaxel by spaxel estimated using the multi-SSP model analysis performed by the \textsc{Pipe3D} pipeline across the FoV of the corresponding datacube, correcting them for the cosmological distance and dust attenuation \citep[see details in][]{pipe3d_ii}. We adopt a Salpeter Initial Mass Function \citep[IMF,][]{Salpeter55}, for these computations. The typical uncertainty for \Mstar\, in our sample is $\sim$ 0.07 dex \citep{pipe3d_ii}.

In addition to the stellar mass, we explore the correlation of the oxygen abundance with a large set of observational properties derived for each galaxy: 
\noindent{

\begin{enumerate}[(i)]

    \item{The integrated molecular mass using as proxy the optical extinction (\MmolExt). We determine \MmolExt\ in a similar fashion as \Mstar, this time integrating the molecular gas surface density \citep[see details in][]{BB20}.}

    \item{The molecular gas fraction from the integrated molecular mass (\FmolExt). We derive the molecular gas fraction as \mbox{\FmolExt = \MmolExt $/$(\MmolExt$+$\Mstar)}. Note that for a sub-sample of CALIFA galaxies we have direct measurements of molecular mass from CO observations for which the molecular gas fraction is defined as \mbox{\FmolInf = \MmolInf $/$(\MmolInf$+$\Mstar)} (see Sec.~\ref{sec:EDGE-CALIFA})}

    \item{Total gas fraction, (\FgastotE). From the estimation of the atomic gas mass, \MHI (See Secc.\ref{sec:HI}) we derive the total gas fraction using both the dust-to-gas proxy, \FgastotE=(\MHI+\MmolExt)/(\MHI+\MmolExt+\Mstar) ) for all galaxies with HI observations, and the direct estimation of the molecular gas, \FgastotI=(\MHI+\MmolInf)/(\MHI+\MmolInf+\Mstar)), for the sub-sample of them with CO observations (156 galaxies).}

    \item{The integrated star formation rate (SFR), derived from dust-corrected \ha\ luminosity, integrated across the FoV of the IFU. Similar to \Mstar, we adopt a Salpeter IMF, and the calibration proposed by \citet{Kennicutt89}.}

    \item{The specific star-formation rate (\sSFR). Using the SFR we derive the specific SFR, defined as the star formation rate weighted by stellar mass, \mbox{sSFR $=$ SFR $/$ \Mstar}.}

    \item{The star-formation efficiency, defined as \SFEExt\mbox{$=$ SFR $/$ \MmolExt} for extinction based data and \SFEInf\mbox{$=$ SFR $/$ \MmolInf} for direct measuments inferred from CO data.}

    \item{The morphological type for each galaxy, which is determined from multiple visual inspections to the available photometric images \citep[see ][]{EspinosaP20, Lacerda20}};

    \item{The SFR derived from the SSP fitting, \SFRssp \citep[][]{GonzalezDelgado16}. This is an emission-line independent measurement of the SFR, derived by estimating the amount of mass formed in the last 32 Myr based on the stellar analysis performed by \textsc{Pipe3D} divided by this time range \citep[See][]{GonzalezDelgado16, Sanchez20}. This tracer presents a good correlation with the SFR derived based on H$\alpha$ \citep[see e.g.,][]{Sanchez2019SDSSIVM,BB21};} 

    \item{The concentration index, (\Cindex), which is obtained from the ratio between the radius enclosing 90 and 50\% of the flux in g-band image.}
    
    \item{The axis ratio of semi-axes, (\ba), derived using an isophotal analysis on the available g-band images \citep{LopezCoba19}.}

    \item{The inclination, (\incli), obtained by projecting semi-axes of an ellipse fitted to the optical image \citep{LopezCoba19}. }
    
    \item{The luminosity distance (\DL) derived from the redshift (\redshiftz) of the object using the adopted cosmology. }

    \item{The cosmological time in which the galaxy reached 50\%, 80\% and 90\% of its actual stellar mass, derived from the analysis performed by the \textsc{Pipe3D} pipeline (\Tfifty,\Teighty,\Tninety, and their corresponding ratios \Tefty\ and \Tnfty). These parameters are derived from the SSP fitting, tracing the shape of the star-formation history in galaxies \citep[e.g.][]{2017A&A...608A..27G}.}

    \item{The equivalent width of \ha\ at 1\Reff\ (\EWre), derived by dividing the estimated \ha\ flux by the density flux of the continuum estimated at the same wavelength.}

    \item{The specific angular momentum at 1\Reff\ (\lambdaRe) which is derived following the prescriptions indicated in \citet{Emsellem2007,2007MNRAS.379..418C}. Derived using PIPE3D, this parameter involves a 2D velocity field as a proxy of the projected angular momentum per mass { and provides} a kinematic classification into slow and fast rotator core systems.}

    \item{The dynamic mass (\Mdyn), calculated from the S$_K$ kinematic parameter (also known as combined velocity \textit{S}cale that includes a constant \textit{K}), which takes into account the ordered { and random} movements \citep[through the maximum rotation speed and through the average speed dispersion in the aperture considered respectively, ][]{2018MNRAS.479.2133A}. This parameter, first introduced by \citet{2006ApJ...653.1027W},  provides a common scaling relation for both early-type galaxies and late-type galaxies.}

    \item{The ionization parameter (\ioniz), defined as the ratio between ionizing photon density and hydrogen density, estimated using the ratio of \otwo $\lambda$3727 versus \othree $\lambda$5007 \citep{morisset16}.} 

    \item{The distance to the Star Formation Main Sequence (SFMS) of the galaxies (\SFMS), defined as the subtraction between the observed and expected SFR for each galaxy if the SFMS were perfectly followed based on its stellar mass (using the relation derived in \citealt{2016ApJ...821L..26C}).}

\end{enumerate}
 }

}

All together, we collect a total of 29 galaxy properties for each galaxy in order to explore whether each of those could have a significant impact in shaping  the oxygen abundance in our sample of star-forming eCALIFA galaxies.

In order to simplify our notation, hereafter, we use the labels \Teighty, \Tefty, \Tninety, SFR, \FmolExt(\FmolInf), \Tnfty, \sSFR, \EWre, \SFRssp, \ioniz, \SFEExt (\SFEInf), \Tfifty, and \DL \ for the logarithmic form of the respective physical parameters.


\begin{figure*}
\centering
	\includegraphics[width=\textwidth]{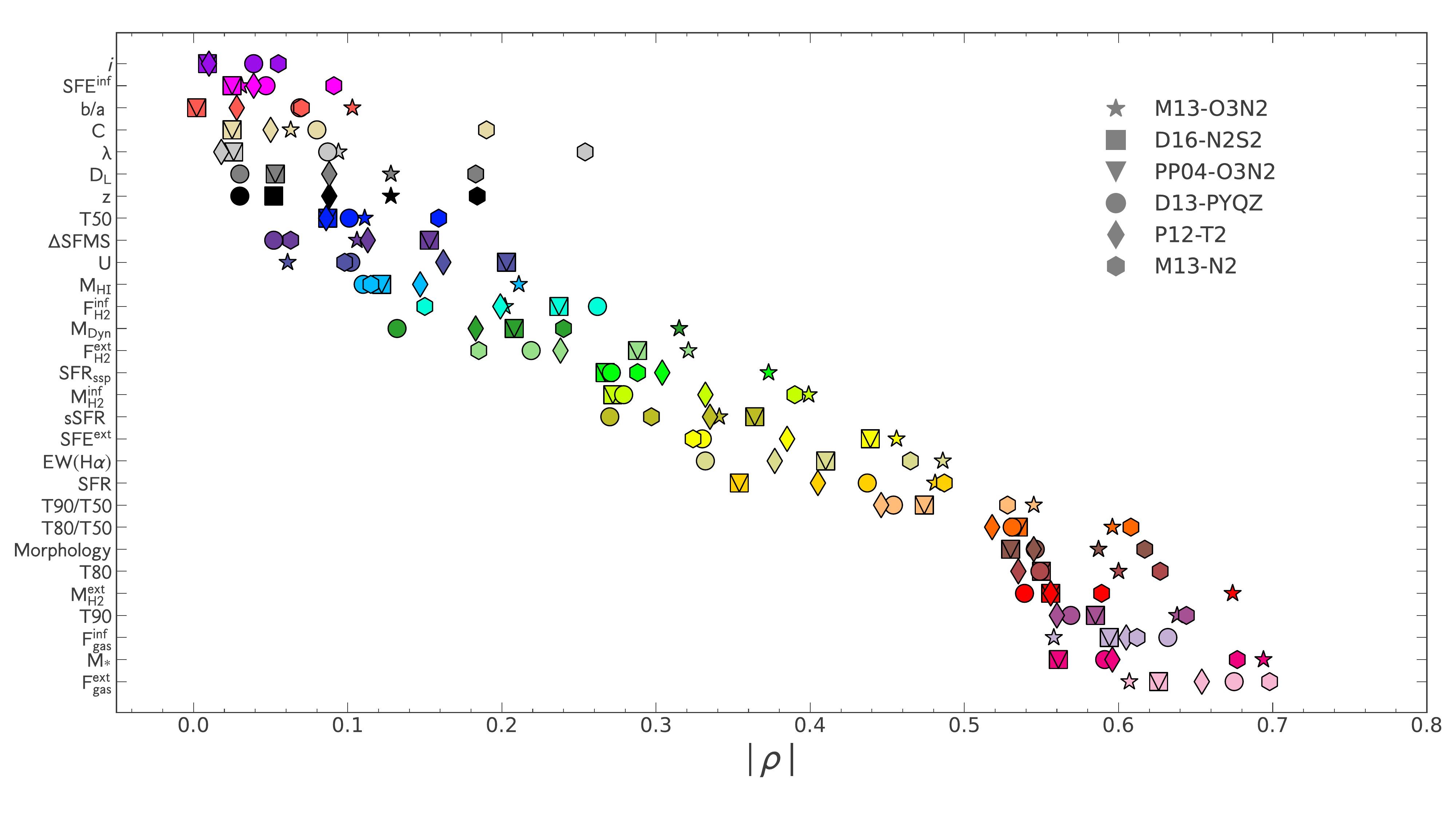}
	\caption{Correlation ranking of our set of galaxy parameters sorted according to { their Spearman's} correlation coefficient, in absolute values ($|\rho|$), with the oxygen abundance (parameters described in Sec.~\ref{sec:oxy_abu}). Each symbol corresponds to each of the six abundance calibrators employed in this study. Each color represents the 29 physical parameters explored here. See legend in the figure. { The total gas fraction, \FgastotE,} is the parameter that correlates the most with the oxygen abundance; other parameters also exhibit large correlation coefficients regardless the calibrator.
	}
\label{fig:cal_all}
\end{figure*}

The complete eCALIFA survey spans a wide range of morphological types and stellar masses ($7<\mathrm{log(M_{\ast}/M_{\odot})}<12$). Its sample and selection { criteria enables a meaningful} analysis for the galaxy population in the nearby universe \citep[e.g.,][]{Sanchez12,walcher14}. We apply { two criteria} to select our sub-sample of analyzed objects extracted from the 906 eCALIFA galaxies: (i) \EWre $>6$\AA. Since our flux measurements are derived at one effective radius, we consider that all our metallicity estimations are indeed corresponding to star-forming regions \citep[i.e., the only ionizing source { is a young} stellar emission,][]{Sanchez13,EspinosaP20}. We corroborate this by plotting the \othree/\hb \ vs \ntwo/\ha \ flux ratios in a BPT diagram \citep{BPT}. We confirm that all the ratios from our selected sample are below the \citet{Kewley01} demarcation line; (ii) $i< 70^{\circ}$. This ensures that our metallicity derivations, estimated at the Re, are not suffering from the different biases introduced by high inclination \citep[e.g.,][]{Ibarra-Medel19}. Based on the above selection criteria, our final sample consists of 299 galaxies. These objects cover a mass range between $7.73<\mathrm{log(M_{\ast}/M_{\odot}})<11.76$ (see Fig.~\ref{fig:logM_Morph}) and also cover a range of oxygen abundance between 8.11$<12+\mathrm{log(O/H)}<$ 8.63.

\subsection{The CARMA-APEX-CALIFA survey}
\label{sec:EDGE-CALIFA}

As indicated before, for a sub-sample of the CALIFA galaxies, we have observations of the CO molecular emission line allowing us to have direct measurements of \Mmol. We use the compilation performed by the Extragalactic Database for Galaxy Evolution survey \citep[EDGE,][]{Bolatto17}, that comprises spatial resolved CO maps on a subsample of 126 galaxies observed with the CARMA array (CARMA: Combined Array for Millimeter-wave Astronomy \citet{2006SPIE.6267E..13B}). In addition we use a dataset of galaxies observed with the Atacama Pathfinder 12m sub-millimeter telescope \citep[APEX,][]{2006A&A...454L..13G}. The APEX dataset was already presented in \citet[][]{APEX} and \citet{Sanchez21}. It comprises a total of 418 estimations of the molecular gas within approximately 1\Reff \ of CALIFA galaxies based on CO emission for an aperture of $\sim$26$\arcsec$.

\subsection{\hi\, measurements}
\label{sec:HI}

In addition to the molecular gas content, the total neutral gas requires to estimate the atomic gas, derived through direct observations of the 21~cm emission line. Similarly to the molecular gas, we do not have access to this information for the full eCALIFA sample. Furthermore, we do not have { either a proxy or a homogeneous} sub-sample of \hi\, observations to derive such a proxy. As an estimation of the total \hi\, gas mass we rely on the compilation of radio observations performed by L\'opez-S\'anchez et al. (in prep.), that comprises a total of 406 integrated values. We should note that since they are based on both single-dish and { interferometric array} observations, there is no homogeneous aperture over which this quantity is derived. However, { in general, the HI extent} much larger than the region over which the optical data or the molecular gas content were derived. This situation is not different than the { one encountered by} other explorations similar to the one proposed here \citep[e.g.][]{2013MNRAS.433.1425B,Hughes13,brown18,2020MNRAS.496..111Z}. Finally, we recover 167 eCALIFA galaxies that satisfy the selection criteria for star-forming galaxies mentioned early.

\section{Analysis and Results}
\label{ref:Results}


\begin{table*}
\centering
\caption{{ Spearman} correlation coefficient ($\rho$) between each galaxy parameter and the oxygen abundance based on different calibrators\label{Tab:Spearman_calib_direc}. We include the parameters derived from the observed data in the respective columns.}
\begin{tabular}{cccccccccc}
\hline
\hline
\\ 
Calibrator & \FgastotE\ (\FgastotI) & \Mstar & \Tninety & \MmolExt\ (\MmolInf)& \Teighty & \Morph  & \Tefty & \Tnfty & SFR \\ 
\multicolumn{10}{c}{}  \\\hline
D16-N2S2 & -0.607 (-0.558) & 0.694 & 0.638 & 0.674 (0.399) & 0.600 & -0.587 & 0.596 & 0.545 & 0.481 \\
M13-O3N2 & -0.626 (-0.594) & 0.561 & 0.585 & 0.556 (0.272) & 0.550 & -0.530 & 0.535 & 0.474 & 0.354 \\
PP04-O3N2 & -0.626 (-0.594) & 0.561 & 0.585 & 0.556 (0.272) & 0.550 & -0.530 & 0.535 & 0.474 & 0.354 \\
D13-PYQZ & -0.675 (-0.632) & 0.591 & 0.569 & 0.539 (0.279) & 0.549 & -0.546 & 0.531 & 0.454 & 0.437 \\
P12-T2 & -0.654 (-0.605) & 0.596 & 0.560 & 0.556 (0.332) & 0.535 & -0.545 & 0.518 & 0.446 & 0.405 \\
M13-N2 & -0.698 (-0.612) & 0.677 & 0.644 & 0.589 (0.390) & 0.627 & -0.617 & 0.608 & 0.528 & 0.487 \\\hline
Mean & -0.648 (-0.599) & 0.613 & 0.597 & 0.578 (0.324) & 0.569 & -0.559 & 0.554 & 0.487 & 0.42 \\\hline
\\
Calibrator & \EWre & \SFEExt (\SFEInf) & \sSFR & \SFRssp & \FmolExt\ (\FmolInf)& \Mdyn & \MHI & \ioniz & \SFMS \\ 
\multicolumn{10}{c}{}  \\\hline
D16-N2S2 & -0.486 & -0.456 (-0.103) & -0.341 & 0.373 & 0.321 (-0.202) & 0.315 & 0.211 & -0.061 & -0.106 \\
M13-O3N2 & -0.410 & -0.439 (-0.002) & -0.364 & 0.267 & 0.288 (-0.237) & 0.208 & 0.122 & -0.203 & -0.153 \\
PP04-O3N2 & -0.410 & -0.439 (-0.002) & -0.364 & 0.267 & 0.288 (-0.237) & 0.208 & 0.122 & -0.203 & -0.153 \\
D13-PYQZ & -0.332 & -0.330 (0.069) & -0.270 & 0.271 & 0.219 (-0.262) & 0.132 & 0.110 & -0.102 & -0.052 \\
P12-T2 & -0.377 & -0.385 (0.028) & -0.335 & 0.304 & 0.238 (-0.199) & 0.183 & 0.147 & -0.162 & -0.113 \\
M13-N2 & -0.465 & -0.324 (-0.070) & -0.297 & 0.288 & 0.185 (-0.150) & 0.240 & 0.115 & -0.098 & -0.063 \\ \hline
Mean & -0.413 & -0.396 (-0.013) & -0.328 & 0.295 & 0.256 (-0.214) & 0.214 & 0.138 & -0.138 & -0.107 \\ \hline
\\
Calibrator  & \Tfifty & \redshiftz & \DL & \lambdaRe & \Cindex & \ba & \incli & & \\ 
\multicolumn{10}{c}{} \\\hline
D16-N2S2 & 0.111 & 0.128 & 0.128 & -0.094 & 0.063 & 0.031 & -0.009 & &  \\
M13-O3N2 & 0.087 & 0.052 & 0.053 & -0.026 & -0.025 & 0.025 & -0.009 & &  \\
PP04-O3N2 & 0.087 & 0.052 & 0.053 & -0.026 & -0.025 & 0.025 & -0.009 & &  \\
D13-PYQZ & 0.101 & 0.030 & 0.030 & -0.087 & 0.080 & 0.047 & 0.039 & &  \\
P12-T2 & 0.086 & 0.088 & 0.088 & -0.018 & 0.050 & 0.039 & 0.010 & &  \\
M13-N2 & 0.159 & 0.184 & 0.183 & -0.254 & 0.190 & 0.091 & 0.055 & &  \\ \hline
Mean & 0.105 & 0.089 & 0.089 & -0.084 & 0.056 & 0.043 & 0.013 & &  \\ \hline
\end{tabular}
\end{table*} 


\subsection{Correlation with galaxy parameters}
\label{sec:corr_coefs}

It is usually assumed that the stellar mass is the best parameter to describe the characteristic oxygen abundance for each galaxy. However, different studies have also explored the correlation between the oxygen abundance and other physical parameters \citep[such as the SFR, molecular mass gas, { etc.,}][]{1992MNRAS.259..121V,Zaritsky94,PilyuginFerrini00,Garnett02,Tremonti04,Erb06,Ellison08,2016A&A...595A..48B,2019MNRAS.484.5587T}. Given the large set of physical parameters derived from both the stellar and ionized { gas components
} for our sample of eCALIFA galaxies (see details in Sec.~\ref{sec:data}), we test whether there is any parameter that could correlate strongly with the oxygen abundance other than the stellar mass. For doing so we estimate the correlation coefficient\footnote{Calculated using NumPy package from \cite{harris2020numpy}} between the oxygen abundance derived using the six calibrators with our final set of 29 galaxy parameters (including stellar mass),  described in Sec. \ref{sec:data}.

\begin{figure*}[htp]
\centering
\includegraphics[width=0.95\textwidth]{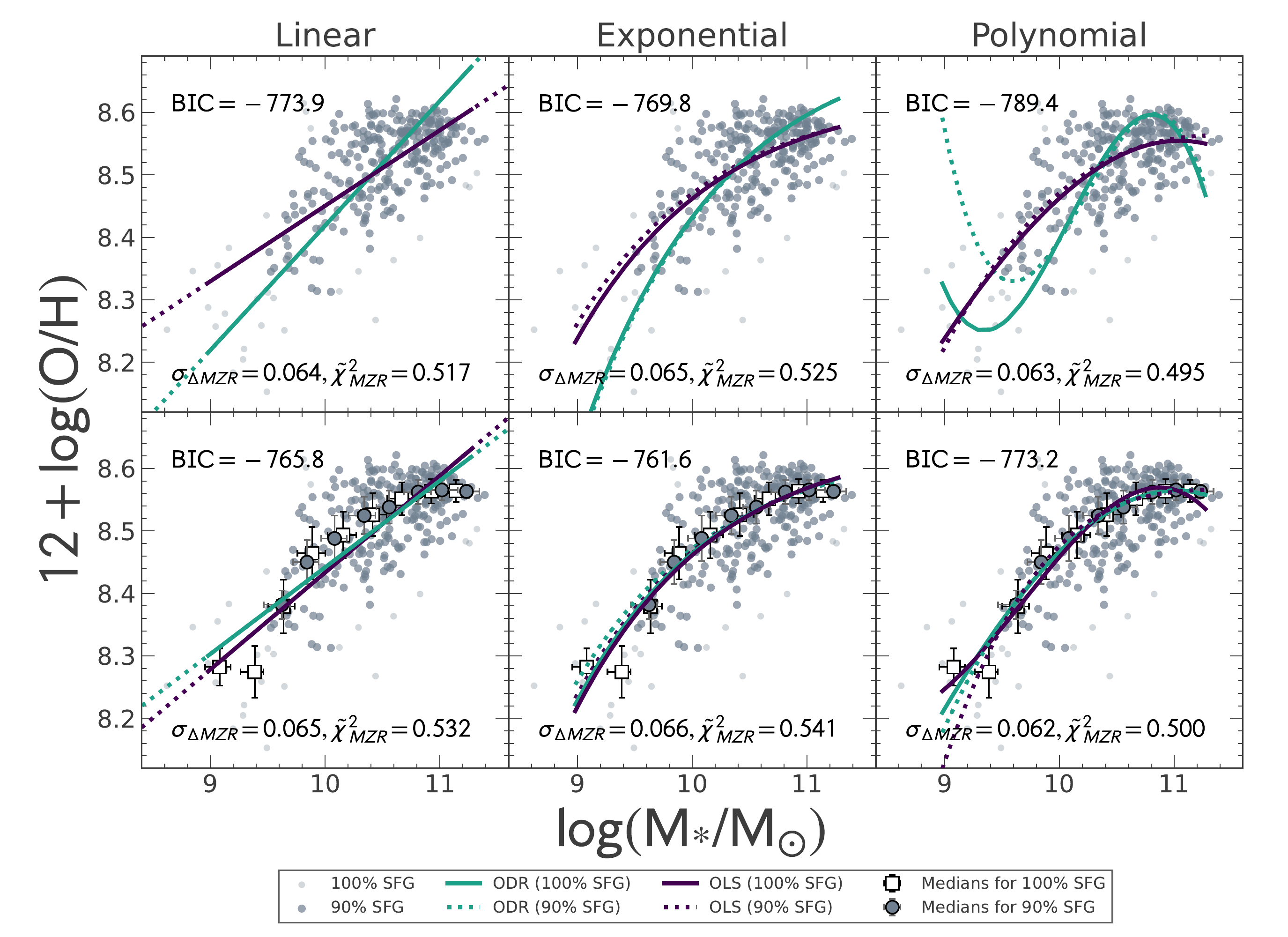}
    \caption{
    Different characterizations of the mass-metallicity relation derived for the { M13-O3N2} calibrator once fitted different functional forms (represented in each column) to our sample of 299 of SFGs (shown as grey circles in all the panels). Upper panels correspond to the derivations fitting the original distribution of points, while bottom panels correspond to the derivations fitting a binned version of the data. In all panels the solid (dotted) purple and green lines represent an OLS and ODR fitting technique for 100\% (90\%) of the sample, respectively. In the bottom panels, the filled-gray circles and the filled-white squares with error bars represent the average metallicity for a given stellar mass bin for 90\% and 100\% of the sample, respectively. The error bar in each of the symbols represents, in the $x$-axis, the size of the stellar mass bin whereas in the $y$-axis, the standard deviation of the binned metallicity. In each panel, $\varrho_{\Delta{MZR}}$ and $\tilde\chi^2_{MZR}$ represent the standard deviation of the residuals and the goodness of the fit (from the residuals against \Mstar). \textit{Similar figure for the rest of the calibrators is found in \url{http://ifs.astroscu.unam.mx/CALIFA/MZR/} }
    }
    \label{fig:m13B1}
\end{figure*}

Fig.~\ref{fig:cal_all} summarizes the result of this analysis, it shows the ranking ordered distribution of the 29 derived correlation coefficients. The values are listed in Table~\ref{Tab:Spearman_calib_direc}. In order to explore for possible secondary relations, we select those parameters that could be associated with significant values of Spearman's coefficients (i.e., $|\rho|>$0.3). Only a total of 14 parameters present a significant correlation with the oxygen abundance, with the gas fraction (\FgastotE) and the total stellar mass being the ones with the strongest correlation. 
The ranking is followed by parameters not directly related with neither the stellar mass or the gas content. It is particularly interesting that the strength of the correlation with \Tninety,  \Teighty,  and \Tefty, parameters that trace directly the shape of the star-formation history in this galaxies  \citep[e.g.,][]{2017A&A...608A..27G}. Together with the morphology, these star-formation history parameters present a stronger correlation, in agreement with \citet{Camps21}, than other parameters more frequently explored as driver/proxies of the oxygen abundance, such as the SFR and the \sSFR. It is remarkable the lack of significant correlations between the oxygen abundance and certain parameters that have been frequently claimed to have a significant impact in the chemical enrichment in galaxies \citep[e.g., \SFE, and \SFMS,][]{Rossi06,CidFernandez07,Tassis08,Panther08,Calura09,salim14}. It is { worth noting} that these results are broadly independent of the adopted calibrator. There is some degree of fluctuation in the ranking order, in particular for those parameters with low values of $\rho$, and a small variation in the actual value for a particular parameter from calibrator to calibrator. However, the main ranking and actual value of the correlation coefficient are essentially the same.

This exploration confirms that \Mstar\ is { one of} the best tracer of the oxygen abundance, at least for our sample and dataset. However, we find that other parameters present a significant correlation, and they may present secondary relations with the abundance. Some of those parameters present strong correlations with the stellar mass as well \citep[e.g., gas fraction, or the SFMS that correlates the SFR][]{2004MNRAS.351.1151B}, and therefore, their real effect on the abundance should be explored { after removing the primary} (i.e., the strongest) relation with the stellar mass. For { doing so, it} is important to determine the relation between the oxygen abundance and the stellar mass in the best possible way, prior to exploring any possible secondary dependency. Since, so far, our results seem to be broadly independent of the adopted calibrator, from now on we focus our analysis using the fiducial calibrator M13-O3N2. We report the results for the rest of the calibrators when required.

\subsection{The shape { of the} mass-metallicity relation}
\label{sec:MZR}

The distribution of the M13-O3N2 oxygen abundance as a function of the stellar masses for our sample of galaxies is shown in each panel of Figure \ref{fig:m13B1}. As in previous studies \citep[e.g.,][among others]{Tremonti04,Ellison08}, we find a rise of the oxygen abundance with the stellar mass, with a possible plateau of asymptotic value reached at the high-mass end (\Mstar$\sim 10^{11}M_{\odot}$). Different procedures have been adopted to parametrize the detailed relation between these two parameters. For instance, in the { literature, it is} common to derive the best fit of the MZR using the average gas-phase metallicity for bins of \Mstar \citep[e.g.,][]{Tremonti04,Ellison08,Zahid14,Tran18,Sanchez16,BB17,Curti20}. However, this type of binning may induce an under-representation of the number of galaxies for a given \Mstar\ in less populated regimes or could be affected by outliers. Therefore, it may be suitable to adopt no binning at all. To explore the impact of the binning, we carry out fits using two different strategies: (i) a no binning scheme (top panels of Fig.~\ref{fig:m13B1}); (ii) stellar mass bins of 0.25 dex width (bottom panels of Fig.~\ref{fig:m13B1}).

We explore the impact of excluding outliers from the data. In this regard, we select those galaxies enclosed by the contour corresponding to 90\% of the objects (i.e., excluding the remaining 10\% as outliers). This reduces the scatter in the distribution, without having a significant impact in the sampled ranges of masses and oxygen abundances.


\begin{deluxetable*}{lcccccccccc}
\tablecaption{Parameters derived for each functional form adopted to characterize the MZR once fitted to the data shown in Fig. \ref{fig:m13B1} .\label{BestFit_paramM13}}
\tablewidth{0pt}
\tabletypesize{\scriptsize}
\tablehead{
\colhead{Fit} & \colhead{Technique} & \colhead{a} &  \colhead{b} & \colhead{c} & \colhead{d}   & \colhead{BIC} &  \colhead{$\tilde\chi^{2}_{MZR}$} & \colhead{$\varrho_{\Delta{MZR}}$}   & \colhead{$\sigma_{\Delta{MZR}}$}& \colhead{$\Delta{\sigma}/ \sigma$}\\ 
\colhead{} & \colhead{} & \colhead{} &\colhead{} & \colhead{} & 
\colhead{} & \colhead{} &\colhead{}  & \colhead{} & \colhead{(dex)} & \colhead{(\%)} 
} 
\startdata
Linear & OLS & 7.24 $\pm$ 0.1 & 0.12 $\pm$ 0.1 & $-$ & $-$ & -773.9 & 0.517 & -0.129 & 0.064 & 29.0 \\
Exponential & OLS & 8.63 $\pm$ 0.1 & -17.57 $\pm$ 0.4 & 3.5 & $-$ & -769.8 & 0.525 & -0.1 & 0.065 & 28.0 \\
Polynomial & OLS & 8.72 $\pm$ 10.2 & -1.07 $\pm$ 3.0 & 0.19 $\pm$ 0.3 & -0.01 $\pm$ 0.1 & -789.4 & 0.495 & -0.043 & 0.063 & 30.0 \\
\enddata
\end{deluxetable*}

In Fig. \ref{fig:m13B1} the white (grey) filled symbols in the bottom panels represent the median values of mass and metallicity in each bin, when using the complete (90\% of the
) the sample. Horizontal error bars represent the size of the stellar mass bin for both cases, whereas the vertical error bars represent the standard deviation of oxygen abundance for each bin. According to this, we explore three functional forms: a linear relation (left panels in Fig. \ref{fig:m13B1}), an exponential relation (middle panels in Fig. \ref{fig:m13B1}) and a third-order polynomial functional form (right panels in Fig. \ref{fig:m13B1}).\\
\textit{(i) Linear form}. Despite the evidence for a saturation in the oxygen abundance at the most massive end \citep[e.g.,][]{Tremonti04, Curti20}, we have included this functional for as the more simple one, and considering that this saturation may be a possible effect of the selected calibrator (Appendix \ref{app_calib}). This functional form is motivated from the observed increasing trend of the oxygen abundance with stellar mass. Considering $y=12+\mathrm{log(O/H)}$ and $x=\mathrm{log(M_{\ast}/M_{\odot}})$, its functional form is:
\begin{equation}
  \label{eq:linear}
  y =  a + bx,  
\end{equation}
where $a$ and $b$ represent the intercept and slope of the relation, respectively.

\textit{(ii) Exponential form}. This functional form naturally describes the flattening in metallicity for massive galaxies reported in the literature \citep[e.g.,][]{Tremonti04,Mannucci10,Zahid14}. Following \citet{Sanchez13}, we adopt the functional form (where $x$ and $y$ corresponds to the same parameters described in Eq. \ref{eq:linear}):
\begin{equation}
  \label{eq:expo}
  y = a+b(x_{8}-c)exp^{-(x_{8}-c))},
\end{equation}
where $x_{8} = x - 8$. We fix $c=3.5$ for our sample, following \citet{Sanchez17}. In this case, the parameters $a$, $b$ and $c$ represent the asymptotic value of oxygen abundance at high masses, the curvature and the stellar mass at which the metallicity reaches its asymptotic value, respectively \citep[see more details in ][]{Sanchez16}.

\textit{(iii) 3rd-order polynomial}. In contrast to the exponential form, this functional form is prone to describe a flattening of the MZR observed at both the high and low-mass regimes, observed by some explorations \citep[e.g., ][]{Tran18,Blanc_2019,Sanchez19}. Despite of the existence of this low-mass flattening, this functional form has been previously adopted in the literature  \citep[e.g.,][]{Tremonti04,Kewley2008,Hughes13,Tran18}.
\begin{equation}
  \label{eq:three}
  y = a+bx+cx^{2}+dx^{3},
\end{equation}
where $x$ and $y$ correspond to the same parameters described in Eq.\ref{eq:linear}. To perform the fit to the data (both binned and unbinned), we use two SciPy fitting techniques \citep[See][]{2020SciPy}, an Ordinary Least Squares (OLS) and an Orthogonal Distance Regression (ODR). Fits based on the OLS technique minimize the sum of square differences of the residuals (i.e., between the observed and predicted values).
In contrast, the ODR technique  consider minimizes the sum of squared orthogonal differences (i.e., perpendicular distances) from the data values.

Both techniques could produce different results and could answer different scientific questions. In other words, the OLS technique explicitly answers how the characteristic oxygen abundance depends on \Mstar, whereas the ODR one tries to understand how are these two parameters related to each other. Therefore, we are interested in studying the physical implications of each framework established by each fitting technique.

The results of this fitting procedure are shown in Fig.~\ref{fig:m13B1}. In general, we find the same trend reported previously in the literature for the MZR \citep{Tremonti04,Ellison08}. Metallicity increases with stellar mass. Indeed, there is a clear correlation between both parameters ($\rho=0.7$). We also observe a possible flattening in the distribution of oxygen abundances at the high-mass end. On the other hand, a flattening for low-mass galaxies is less evident \citep[][]{Hughes13,Blanc_2019}. Considering only the 90\% sample, we note that most of the galaxies are massive ($\gtrsim 9.5 \log(M_{\ast}/M_{\odot})$). We further explore the impact of considering the entire sample or its 90\% when performing the binning and the fitting. 

The bottom panels of Fig.~\ref{fig:m13B1} show the binning scheme for the entire and 90\% of the sample (white-squares and gray-circles with errorbars, respectively). The binning of the data enhances the trends previously described. For the total sample we note an increasing trend for intermediate stellar masses ($9.5<\mathrm{log(M_{\ast}/M_{\odot})}<10.6$), and a flattening in both, low and high mass ends of the MZR. As mentioned above, the flattening for low-mass galaxies is less evident when we consider only 90\% of the sample. However, in both samples, the flattening at massive galaxies is evident.

Considering each of the above selection criteria and binning strategies, we perform a total of 24 fits. To estimate the uncertainty of the best-fitted parameters, we use a Monte Carlo simulation of 1000 realizations. For each of these realizations, we derive a new dataset for the 299 galaxies considering the associated uncertainty for each parameter (typically $\sim$ 0.07, and $\sim$0.06 dex, for the stellar mass and the metallicity, respectively). Finally, we apply the full procedure described above for each of these realizations (i.e., sample selection, binning, and fitting).


\begin{figure*}[htp]
\centering
	\includegraphics[width=0.9\textwidth]{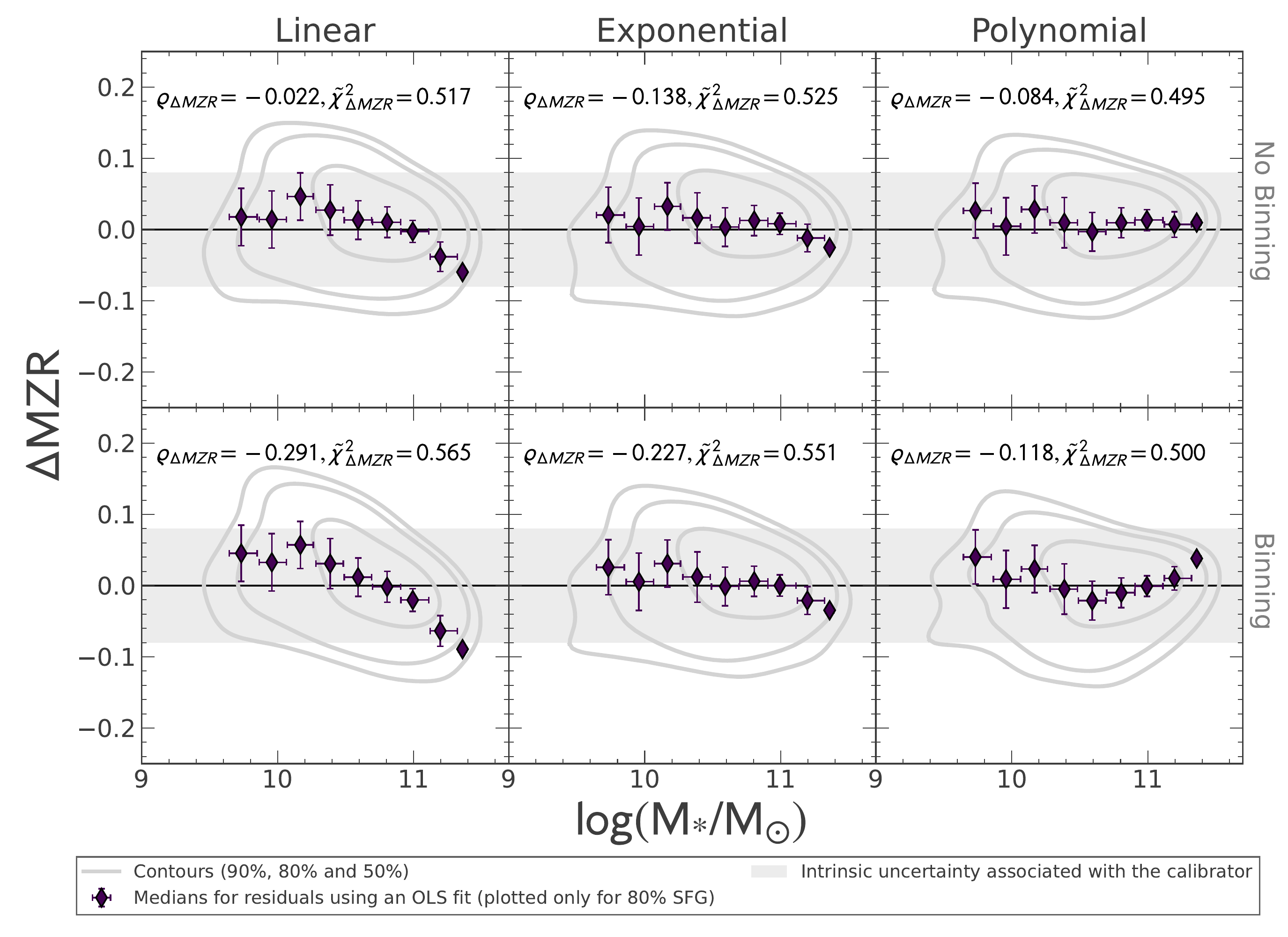}
    \caption{Residuals of the MZR derived using the OLS fitting procedure for the different functional forms (columns) and binning procedures (rows) tested in this study (as shown in Fig. \ref{fig:m13B1}), as a function of the stellar mass. Contours represent the density distribution of $\Delta$MZR, with each consecutive contour encircling 90\%, 80\% and 50\% of the points, respectively. The average values of these residuals in bins of 0.3 dex in stellar mass are shown as solid diamonds, with the errorbars corresponding to the standard deviations around these mean values. The correlation coefficients between the representes parameters and the $\tilde\chi^{2}_{\Delta{MZR}}$ for the linear relation between them are indicated in the legend of each panel. The shaded region indicates the intrinsic uncertainty associated with the { M13-O3N2} calibrator \citep[$\sim$ 0.08 dex,][]{marino13}. 
    }
    \label{fig:m13B2}
\end{figure*}

We show the best fitted models for each combination of functional forms, binning  of the data and fitting procedure in Fig.~\ref{fig:m13B1}. We find that using the entire dataset (i.e., not binned), regardless the adopted functional form, the OLS fit (purple solid lines) yields a significant scatter reduction in comparison to the ODR fit (green solid lines). Furthermore, we do not find significant difference when considering the entire sample or the 90\% (purple and green dashed lines). In other words, the OLS fitting provides a more reliable description of the MZR in comparison to the ODR. This may highlights the significant dependence of the characteristic oxygen abundance with the \Mstar. In addition, we note that, depending on the fitting technique, the shape of the MZR is significantly different when using the polynomial functional form. 

In contrast to the fittings using the whole sample, the ODR and the OLS fits are quite similar using the binning procedure regardless the adopted functional form (see bottom panels of Fig.~\ref{fig:m13B1}). However, qualitatively the functional form that best describes the binned data is the polynomial one. Unlike the results obtained when using the ODR fit for the polynomial functional form and the entire un-binned sample, we do not find a turnover in characteristic oxygen abundance at high stellar mass, suggesting that this may be an spurious artefact of the previous procedure (thus, a pure plateau at high mass is a more general representation of the data). Furthermore, the flattening at the low-mass end is less evident when considering a polynomial fit using the complete sample (comparing the un-binned and binned fits). 
To further quantify the difference between the best fit using different functional forms and binning, we derive for each of them {an estimation of the goodness-of-the-fit measured by the Bayesian Information Criterion \citep[BIC, ][]{Kass_Raftery_1995} and the reduce chi-squared test ($\tilde\chi^{2}_{MZR}$), as well as the standard deviation of the residuals of the oxygen abundance with respect to the best fitted model ($\sigma_{\Delta{MZR}}$). These three values are included in each panel of Fig.~\ref{fig:m13B1}, for the fit that yields the smallest correlation coefficient of the residuals and the smallest BIC of the MZR.
In Table~\ref{BestFit_paramM13} we present the values of the parameters that provide the best fit for each of the functional forms using unbinned data (the fitting parameters for all fitting combinations are listed in Table \ref{FitParam_M13} in Appendix \ref{app:24_fit_param_m13}). In addition, we list the Spearman correlation coefficient of the residuals with the stellar mass ($\varrho_{\Delta{MZR}}$), the goodness-of-fit using the BIC and the reduced chi-squared of each fit ($\tilde\chi^{2}_{MZR}$), the standard deviation of the residual ($\sigma_{\Delta{MZR}}$) and the relative reduction of the standard deviation once subtracted this best fit ($\Delta\sigma / \sigma=(\sigma_{log(O/H)}-\sigma_{\Delta{MZR}}) / \sigma_{log(O/H)}$).}

We perform a similar analysis for the binned data. Although we find similar values for the above parameters, the OLS polynomial fit using the binned data yields slightly larger values of $\tilde\chi^{2}_{MZR}$, and $\Delta\sigma / \sigma$. The three functional forms for the different explored treatment of the data provides with a similar standard deviation of the residual ($\sigma_{\Delta{MZR}}$), being the polynomial fit using the OLS only smaller in the third decimal. This difference is far smaller than the reported systematic uncertainty of the adopted calibrator. If the parameter used to determine the goodness of the fit is $\tilde\chi^{2}_{MZR}$, the analysis suggests that this fit (polynomial form with OLS fit for no binned data) provides the best characterization of the relation. However, the differences are subtle when using the same functional form and any fitting procedure over the binned data. {Using the Bayesian Information Criterion on the unbinned data, we still find that the polynomial fit provides the best description of the MZR (see Table \ref{FitParam_M13}).

Besides minimizing the $\sigma_{\Delta{MZR}}$, BIC or $\tilde\chi^{2}_{MZR}$, the functional form that best describes the MZR should not present a significant correlation with the abscissa (i.e., the stellar mass)}. Figure~\ref{fig:m13B2} shows the distribution of the residuals of the oxygen abundance with respect to the best functional form (\DeltMZR) against the stellar mass for the entire sample (i.e., no binning, top panels) and the binned data (bottom panels). In all cases it was adopted the MZR relation derived using the OLS fitting procedure, based on the results of the analysis described above. For each dataset it was derived the correlation coefficient between \DeltMZR \ and \Mstar, fitting a linear relation between both parameters and estimating the corresponding  ${\tilde{\chi^{2}}}_{\Delta{MZR}}$. A visual comparison of the trends between the residuals and the stellar mass shows that: (1) the residuals from the analysis based on the non-binned data present a weaker correlation with the mass for all functional forms; (2) the distribution of average residuals (solid diamonds) present the weakest apparent dependence with the stellar mass for the MZR derived using the polynomial functional form (i.e., the values present the lowest deviation with respect to the zero-value). Finally, this functional form applied to the MZR derived for the non-binned data provides with the lowest $\tilde\chi^{2}_{\Delta{MZR}}$ and the 2nd lowest $\varrho_{\Delta{MZR}}$. Once again, the differences are subtle for this functional form when comparing the binning and non-binning treatment of the data.

In summary, from our analysis we find that a polynomial of third-order is the functional form that best describes the relation between the characteristic oxygen abundance (measured at one effective radius) and the total stellar mass for the sample of star-forming eCALIFA galaxies. Finally, we suggest that in order to have a reliable shape of the MZR it is required to adopt (i) an OLS procedure for the full data set or (ii) { either an} ODR or an OLS procedure for a binned dataset. From those cases, (i) provides a slightly better characterization of the MZR.

For completeness we provide in  Appendix \ref{app:MZR_calibs} (and Fig. \ref{fig:MZR_calibs}) the best characterization of the MZR relation using the full set of oxygen abundance calibrators explored in this article. We repeated exactly the same procedure adopted to derive the best fit using our fiducial calibrator, M13, for the other calibrators. No significant differences are found for any of the explored calibrators. Thus, the best fit is achieved using the OLS technique, over the un-binned data, for a polynomial functional form, despite the fact that the actual parameters derived for the MZR relation are different.

\begin{figure}
\centering
	\includegraphics[width=\linewidth]{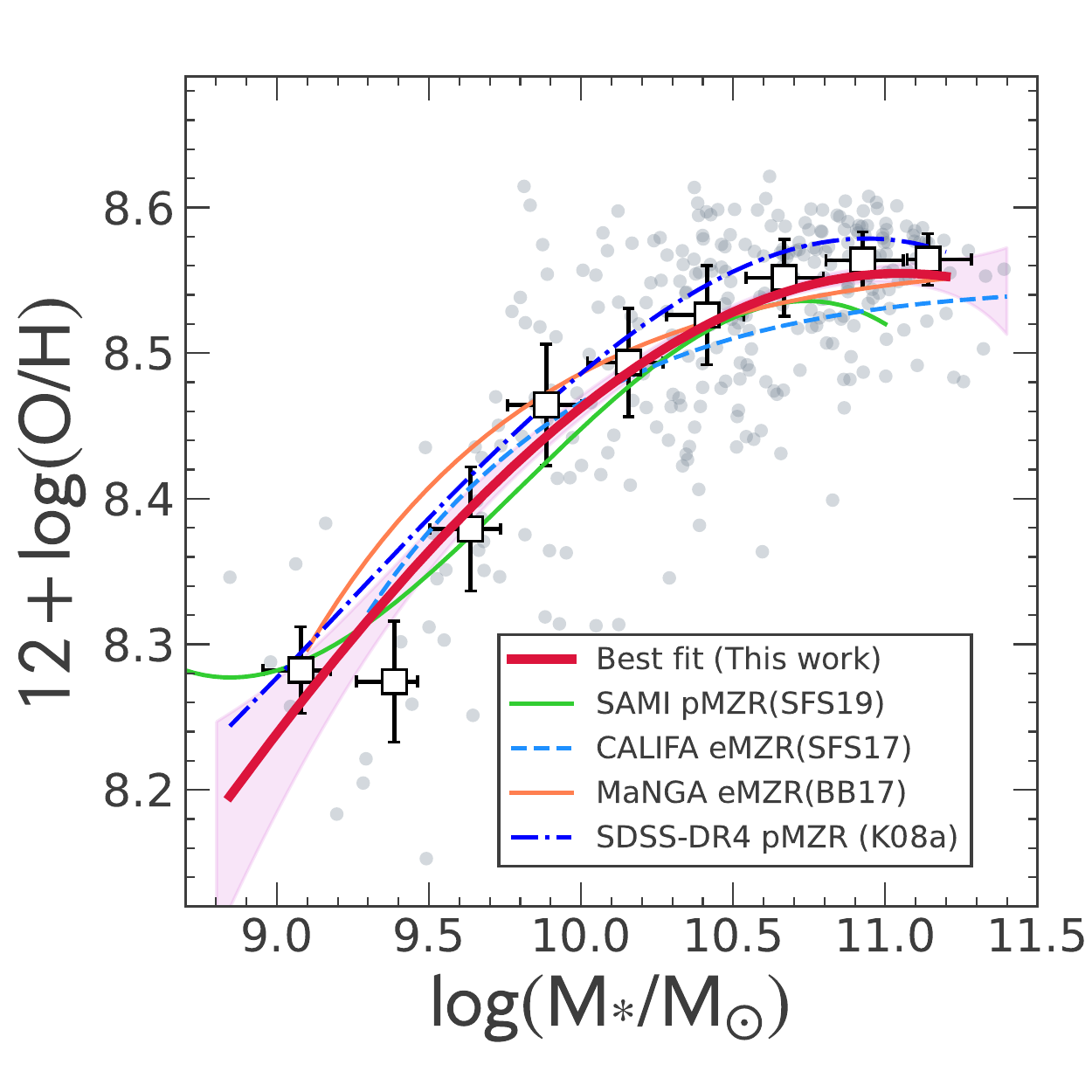}
    \caption{Comparison of the best fitted MZR relation based on our analysis for similar results extracted from the literature. Grey circles represent the distribution along the O/H-\Mstar\ diagram for the eCALIFA star-forming galaxies. White solid squares with error bars represent the median values per bin of stellar mass with errors representing one standard deviation of the oxygen abundance and the size of each bin. The solid-orange, dashed-blue, solid-green, and dash-dotted-purple lines represent the best fit curve for the MZR for the MaNGA \citep{BB17}, CALIFA \citep{Sanchez17}, SAMI \citep{Sanchez19}, and SDSS Data Release 4 \citep{Kewley2008} datasets, respectively. The solid red line is our best fit while the red shaded region represents the range of results derived by the fitting procedure based on a MonteCarlo simulation (see details in Sec.~\ref{sec:MZR}).}
    \label{fig:MZR_comparison}
\end{figure}

\subsection{Comparison with the literature}
\label{sec:lit}

To provide a fair comparison between the shape of the MZR derived in this study and those reported in the literature, we here those studies that follow a similar strategy as ours. Thus, we restrain ourselves to compare our results with other global estimations of the MZR, thus avoiding possible effects of the different adopted procedures (e.g., metallicity calibrator, sampled area, etc.). In Fig.~\ref{fig:MZR_comparison} we present a comparison of the best fit derived in Sec.~\ref{sec:MZR} using our dataset (solid red line) with those extracted from the literature from different { datasets that} measured the oxygen abundance at the same distance (\Reff) and also used the same metallicity calibrator \citep{marino13}. This comparison suggests that regardless of the IFS dataset or survey employed to derived the best fit of the MZR, the derived shape is consistent (within uncertainties) with the one derived in this study. { For polynomial or exponential MZR fits, we designate them as}  pMZR or eMZR, respectively. In \citet{Kewley2008} (here and after, K08), a polynomial shape is proposed { for the SDSS DR4} dataset based on the O3N2 calibrator from \citet{pp04}. We use a linear transformation between the PP04-O3N2 calibrator and the { M13-O3N2} calibrator to include the K08-polynomial fit (K08a; purple continuous line). It yields a slightly shifted polynomial towards larger values of metallicity. Showing a soft low-mass end similar to our MZR. We note that in comparison to \citet{Sanchez17} (SFS17, blue dashed line in  Fig.~\ref{fig:MZR_comparison}) who used the same initial dataset but included highly-inclined galaxies (612 galaxies), our best fit of the MZR is similar to theirs (note they adopted an exponential fit instead of a polynomial one). This may suggest{ the small impact} that inclination may induce in this relation. However, their  underestimation of the oxygen { abundance of massive} galaxies could suggest either the impact of the inclination or the difference in the adopted shape of the MZR. With a larger sample drawn from the MaNGA survey \citep[][$\sim$ 1700 galaxies]{Bundy_2014} using an exponential fit, \citet{BB17} found a similar shape of the MZR as the one reported in this study (BB17, orange continuos line in Fig.~\ref{fig:MZR_comparison}). Finally, using $\sim$ 1000 galaxies from the SAMI survey that observes more galaxies below $10^9$ stellar masses in comparison to studies with CALIFA and MaNGA samples \citep[][]{Croom_2012}, \cite{Sanchez19}, suggested that a polynomial form provides the best description of the MZR, since the SAMI dataset includes more low massive galaxies compared to our sample (see green solid line in Fig.~ \ref{fig:MZR_comparison}). This was motivated, partially, due to the flattening of the MZR at low-mass galaxies. Indeed, for the low-mass regime in our sample, the binned metallicities agree well with their function form. In contrast to the MaNGA or the CALIFA original sample, our sample of galaxies includes a larger number of low-mass galaxies (similar to SAMI). This may reflect the possibility that a large sampling of oxygen abundance at the low-mass regime could indeed reveal a flattening in the MZR at this end \citep[e.g.,][]{Blanc_2019}. In addition, when we explore the variation of the shape of the MZR using a Monte Carlo approach (see red shaded area in Fig.~\ref{fig:MZR_comparison} ), we note that for the associated uncertainties and statistics of our dataset at the low-mass end, we cannot rule out either a flattening or increase of the oxygen abundance with respect to the stellar mass. Explorations based on very low-mass galaxies suggest a monotonic decrease of oxygen abundance \citep[e.g.,][]{2006ApJ...647..970L,2008IAUS..255..397R,2016MNRAS.456.2140M}, indicating that there is no flattening of the MZR at this mass range. Further studies with larger samples of low-mass galaxies using IFS data are required to determine the existence of a low-mass flattening and furthermore to understand the physical scenario that explains it. 

A similar comparison using the other calibrators adopted along this work is available in App.~\ref{app:MZR_calibs} in Fig.~\ref{fig:MZR_calibs}. Although there are subtle differences and in some cases systematic offsets there is a general agreement between our results and those reported in the literature. Once we derive the best functional form and fitting procedure to characterize the MZR relation, we can proceed to explore the dependence of its residuals with respect to other galaxy's parameters that could impact on its shape.

\begin{figure*}
\centering
	\includegraphics[width=0.95\textwidth]{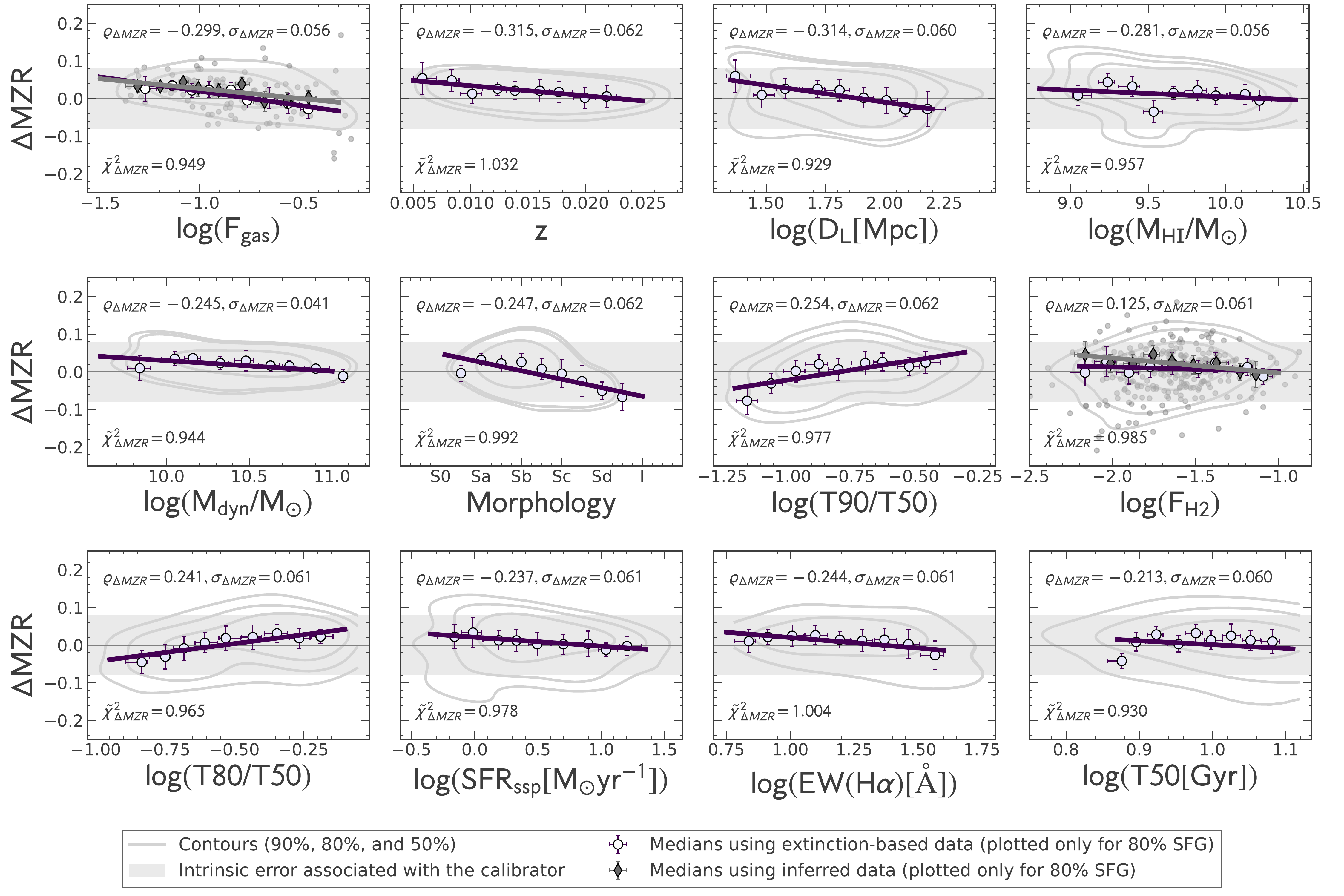}
    \caption{Distribution of the MZR residuals once subtracted the best fitted relation, \DeltMZR, for the { M13-O3N2} calibrator, as a function of the first 12 galaxy parameters ranked in order of the average correlation coefficients. Each panel corresponds, from top-left to bottom-right, to the distributions along { \FgastotE , \redshiftz, \DL, \MHI, \Mdyn, \Morph, \Tnfty, \FmolExt, \Tefty, \SFRssp, \EWre\ and \Tfifty} In each panel, the distribution of \DeltMZR\ is represented by grey density contours (encircling 90\%, 80\% and 50\% of the sample, respectively). Solid white circles represent the average of the \DeltMZR\ within bins of the considered parameter. Vertical error bars correspond to the standard deviation. The solid purple line corresponds to the best fitted linear regression to the white solid-circles. The correlation coefficient between \DeltMZR\ and each parameter and the standard deviation of the residual between \DeltMZR\ and the solid purple line are indicated in the top of each panel, together with the reduced $\tilde{\chi}^2$, at its bottom. The grey shadow region represents the standard deviation of the original residual in each panel ($\sim$ 0.08 dex). For the parameters involving the molecular gas, we include in the figure those galaxies with direct estimations of molecular gas using the EDGE-APEX-CALIFA dataset (\FgastotI\ and \FmolInf). In this case, the average residuals using these data within each bin are represented by solid grey-diamonds and the best fitted linear regression by a solid grey line. 
    In contrast to correlations found in Sec.~\ref{sec:corr_coefs}, \DeltMZR\ shows no significant correlations with other physical parameters (i.e., $\rho<$0.3 in all cases). \textit{Similar figure for the rest of the calibrators is found in \url{http://ifs.astroscu.unam.mx/CALIFA/MZR/}}
    }
    \label{fig:m13B3}
\end{figure*}

\subsection{\DeltMZR\ as a function of galaxies' properties}
\label{sec:Residuals}

So far we have demonstrated that the strongest correlation between the oxygen abundance and different galaxy parameters is the one with the stellar mass (Sec. ~\ref{sec:corr_coefs}). Subsequently, we have derived the best possible characterization of the relation between both parameters, i.e., the MZR relation (Sec.~\ref{sec:MZR}). This best-fitted functional form minimizes the dependence of the residual with the stellar mass, effectively removing any dependence with this parameter (see bottom-right panel of Fig~\ref{fig:m13B2}). On the other hand, previous studies have suggested that the shape of the MZR may be affected by other physical properties of the galaxies, in particular by the SFR \citep{Mannucci10} and the gas fraction \citep[e.g.][]{2013MNRAS.433.1425B}. In other words, they predict a dependence of the residual of the MZR with those additional parameters. Our analysis in Sec. \ref{sec:corr_coefs} indicates that besides the stellar mass there are other 14 physical parameters showing a significant correlation ($\rho > 0.3$) with the oxygen abundance. Some of those parameters are strongly correlated with the stellar mass, and therefore their correlation with the oxygen abundance may be induced by this primary dependence. { Regardless, we consider} that they are the best candidates to produce a possible effect in the { $\Delta{MZR}$; this is} a secondary relation with the oxygen abundance, which in turn { could alter the} shape of the MZR.

\begin{figure*}[htp]
\centering
	\includegraphics[width=1.0\textwidth]{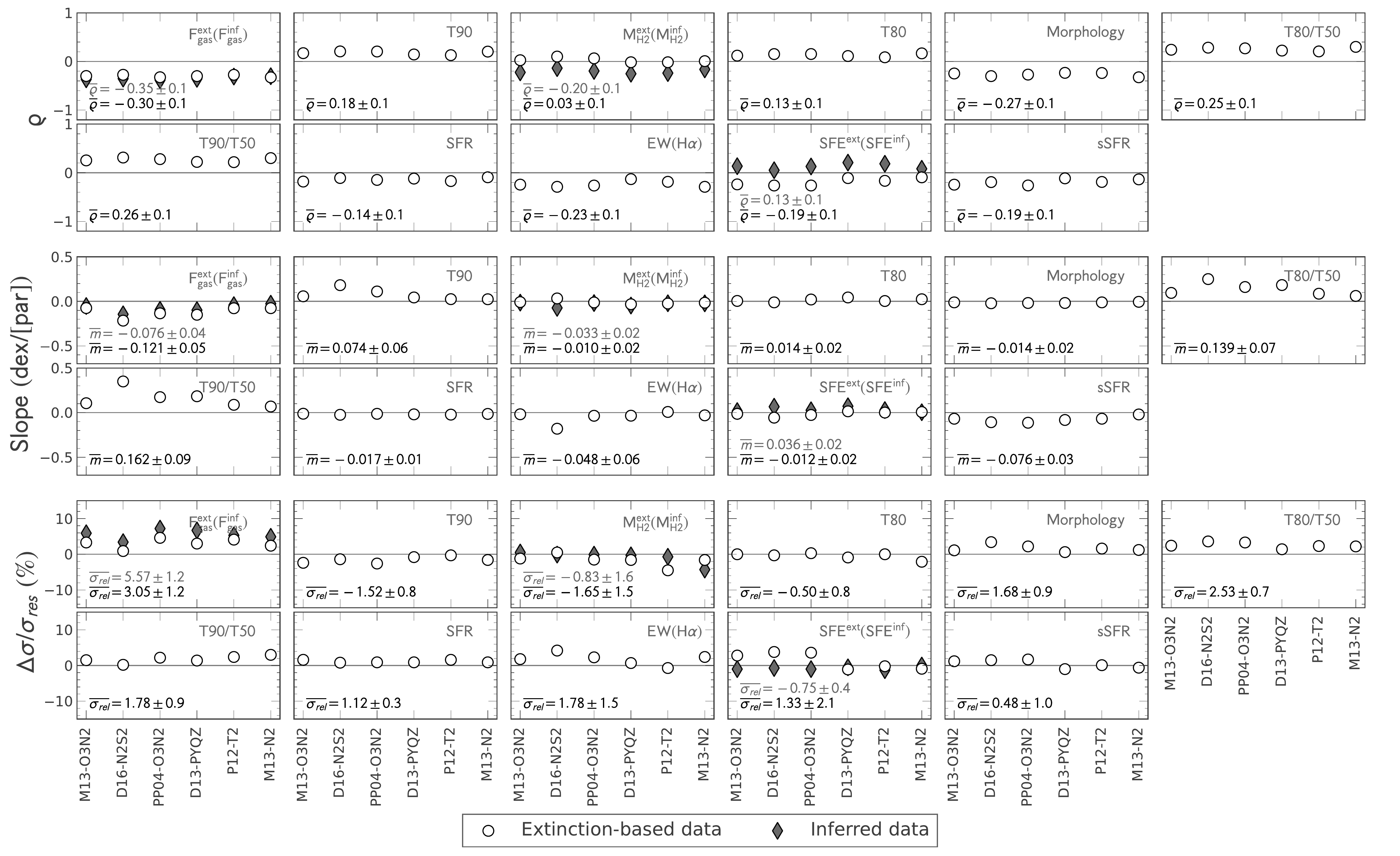}
    \caption{Distribution of the correlation coefficients (upper panels), slopes of the linear relation (middle panel) and relative decreases of the standard deviation (lower panels), of the residuals of the MZR along each of the explored parameters (indicated in the inset label), for each of the six explored oxygen abundance calibrators (x-axis, in each panel). White circles represent the results based on the eCALIFA star-forming sub-sample, while grey diamonds corresponds to the results based on the EDGE-APEX subsample. Each panel in the top, middle, and bottom rows show the values of $\varrho$, $m$, and $\Delta \sigma / \sigma_{res}$, respectively, for the set of calibrators for each physical parameter following the order presented in Fig.~\ref{Tab:Spearman_calib_direc}.
    Mean values and standard deviations of the represented values using the eCALIFA (EDGE-APEX) data are showing in black (grey) inset labels. }
    \label{fig:slopes}
\end{figure*}

In this section we explore if any of those parameters could in fact exhibit a significant correlation with the scatter of the MZR and thus be a secondary parameter { that affects the} metallicity other than the stellar mass. In Fig.~\ref{fig:m13B3} we show the distribution of \DeltMZR\ as a function of the { 12 first parameters that present a mild correlation with the residuals of the oxygen abundance for our sample. We sort these panels according to their average correlation ranking (decreasing $\varrho_{\DeltMZR}$}; from top to bottom, left to right, taking the absolute average values listed in Table \ref{Tab:Spearman_calib_direc}) derived in Sec.~\ref{sec:corr_coefs}. In general, we find that \DeltMZR\ does not show a significant correlation with any of those physical parameters. This is quantified by their Spearman's correlation coefficient, $\varrho$ (see values in each of the panels in Fig.~\ref{fig:m13B3} as well as in the first row of Table~\ref{Tab:BestFit_paramCalib}). The correlation coefficients derived between the \DeltMZR \ and any of the explored parameters are significantly smaller that the correlation coefficients found for the oxygen abundance (see Fig.~\ref{fig:cal_all} and Tab.~\ref{Tab:Spearman_calib_direc}). This suggests that those correlations derived using the oxygen abundance and the physical parameters reflect an intrinsic relation between these parameters and the stellar mass (or among themselves), rather than a particular dependency with the oxygen abundance itself. We also note that a significant fraction of galaxies ($\sim$ 80\%) { show a $\Delta{MZR}$ smaller} than the systematic error associated with the calibrator (gray shaded areas in Fig.~\ref{fig:m13B3}). This is evidence further suggesting that those parameters do not have a significant impact in shaping the oxygen abundance instead of the stellar mass.

Although for most of the physical parameters explored, \DeltMZR\ has a clear lack of dependence (e.g., \Teighty, \Tninety, \SFEExt(\SFEInf), \Cindex, \lambdaRe), there are some that may hint a possible impact in the MZR (e.g., \FgastotE (\FgastotI), \redshiftz (or \DL),\MHI,\Mdyn, \Morph, \Tnfty, \FmolExt(\FmolInf), \Tefty, \SFRssp, \EWre, \Tfifty). To account for this possible dependence we derive the best linear fit of $\Delta{MZR}$ and each of the physical parameters explored in this section (see purple solid lines in each panel of Fig.~\ref{fig:m13B3}). Following the same binning procedure described in Sec.~\ref{sec:MZR}, we derive this fit by using the average values of $\Delta{MZR}$ in bins of each parameter. In order to avoid the impact of outliers, these bins are derived only for those galaxies which values are encircled in the 80\% density contour. The slopes derived from this fitting procedure for our fiducial calibrator are listed in Table~\ref{tab:statistical_parameters}. At first order we assume that any possible relation between the residuals and the considered parameter could be characterize by a simple linear relation. This may be a simplification, but a visual inspection of the distribution of data (and the median values within each bin) indicates that this description should be enough to characterize the observed trends. Furthermore, a linear relation has the advantage that the slope can be used to gauge the strength of the relation between the parameters. We conclude from this exploration that we cannot recover any clear secondary dependence based on strength of the correlation coefficient or the slope of the derived linear relation. 

As we mention above, the relevance of a secondary parameter in the MZR can be tested by comparing the standard deviation of the original residuals ($\sigma_{\Delta{MZR}}$) with those obtained once subtracted the described linear relations (denoted as $\sigma_{res}$). A significant secondary dependence of a given parameter would yield a significant smaller $\sigma_{res}$ in comparison to $\sigma_{\Delta{MZR}}$. We quantify this possible reduction in the scatter by the relative comparison of these two values, i.e., $\Delta{\sigma}/\sigma_{res}$, where $\Delta{\sigma}=(\sigma_{\Delta{MZR}} - \sigma_{res})$. This value is smaller than a 4\% for all the explored parameters. Thus, we do not find significant evidence of a reduction of the scatter of the MZR when introducing a secondary dependence with any of the explored parameters.

We recall that to provide an estimation of the molecular gas for our entire sample of star-forming galaxies, we adopt as a proxy for the molecular gas mass the integrated optical extinction derived from the Balmer decrement \citep[\MmolExt,][]{BB20}. It could be the case that this estimation may introduce a bias with respect to \DeltMZR. Therefore, we also reproduce the comparison of these residuals vs. \MmolInf\ for a sub-sample of galaxies with CO observation (see Sec.~\ref{sec:EDGE-CALIFA}). We find a similar trend and slope as those derived using the extinction as proxy for \MmolInf\ (see gray points and { solid lines in three panels in Fig.}~\ref{fig:m13B3}). Using the direct CO observations we find even a smaller correlation between \FmolInf\ and the residuals of the MZR. As with the other parameters, for the fiducial calibrator we also include the Spearman correlation coefficient, slope, and scatter reduction using both the \MmolInf\ and \FmolInf\ (see Table \ref{tab:statistical_parameters}). Furthermore, we compare the CO-based with the Av-based derivations. We find that the Spearman coefficient and the slope increases (or decreases) for certain calibrators and parameters, specially for \FmolInf. However, all these values remain near to zero.

We repeated the entire analysis outlined above for the full set of calibrators adopted along this article finding very consistent results (see App. \ref{app:D_MZR_calibs}). The results of this analysis are summarized in Fig.~\ref{fig:slopes}, where we represent, for each calibrator the correlation coefficient ($\varrho$) between the $\Delta{MZR}$ and each of the explored parameters, together with the slope and relative reduction of the scatter $\Delta \sigma / \sigma_{res}$ once introduced a linear regression between them. In summary, we find that (i) there is no significant correlation, for any of the calibrators, between the residual of the MZR relation and any of the 14 explored parameters; and (ii) the ratio between the observed scatter of \DeltMZR \ and the scatter derived from a linear fit to the \DeltMZR-parameter relation is very similar, suggesting the little impact that those 14 parameters have in shaping the MZR. No significant scatter reduction is appreciated once we introduce a linear fit between \DeltMZR\ and the considered parameter. Finally, the reported results are independent of the adopted calibrator.

\section{Discussion}
\label{sec:Discussion}

\subsection{Does the oxygen abundance depend on a physical parameter other than $M_{*}$?}
\label{disc:corr}

Several studies have suggested that the oxygen abundance depends, beside the stellar mass, on other observables including structural parameters \citep[e.g.,][]{Tremonti04,Wuyts16,Eugenio18,2021ApJ...908..226H}, star formation properties \citep[e.g.,][]{Rossi06,Ellison08,Mouhcine08,Tassis08,Mannucci10,LaraLopez2010,Yates12,Curti20} and gas-phase properties \citep[e.g.,][]{1992MNRAS.259..121V,Garnett02,Brooks_2007,2013MNRAS.433.1425B, 2016MNRAS.456.2140M, 2016A&A...595A..48B,2019MNRAS.484.5587T}. Furthermore, some of these studies have stated the existence of a so-called fundamental metallicity relation (FMR); this is a secondary dependence between the SFR (molecular gas) and the MZR \citep[see review in][]{2019A&ARv..27....3M}. However, studies using large samples of IFS dataset \citep[e.g.,][]{BB17, Sanchez17, Sanchez19} or even using the SDSS dataset \citep[e.g.,][]{2016ApJ...823L..24K,Telford16} have not found a strong correlation between the MZR and the SFR. In contrast, \citet{2019A&A...627A..42C} claimed that the oxygen abundance depends on the SFR at fixed mass, based on their re-analysis of the same IFS datasets, arguing that their results are consistent with a secondary dependency with the SFR.

In Sec.~\ref{sec:corr_coefs} we find that indeed the stellar mass is the parameter that best correlates with the oxygen abundance in star-forming galaxies. We also find that there are other physical parameters exhibiting a significant correlation with the metallicity, although smaller than \Mstar\, (e.g., \Fgastot, T80, and the morphology, but not strongly on the SFR; see Fig.~\ref{fig:cal_all}). This analysis suggests that \Mstar\, can be considered as the best physical parameter that describes the observed distribution of oxygen abundance out of a wide range of physical parameters. It is also suggesting that potentially there are other galaxy parameters that could have a significant impact in shaping the characteristic metallicity of galaxies in the nearby universe.

Having determined that the best proxy for the oxygen abundance is the stellar mass, in Sec.~\ref{sec:MZR} we proceed to carefully estimate the functional form that best describes the MZR for our dataset. Then, we derive the residuals of the oxygen abundance once subtracted the best fit to this functional form. We observe that subtracting the dependence with the stellar mass produces a significant reduction of the scatter of $\Delta\sigma / \sigma \sim 30\% $ for any of the explored oxygen abundance calibrators. 

Using the SDSS dataset, \cite{Mannucci10} argue that the result of introducing a secondary dependence on the MZR is a significant reduction in its scatter (their Fig. 5), as expected. However, the residuals of the MZR when this secondary relation is not taken into account should correlate with that physical parameter \citep[this has not been considered in more recent explorations, e.g., ][]{2019A&A...627A..42C}. In Sec.~\ref{sec:Residuals}, we use \DeltMZR\ to quantify whether there is another parameter that could have a potential impact in shaping the metallicity other than the stellar mass (see Fig.~\ref{fig:m13B3}). Our analysis shows that although there are some parameters that may exhibit very weak correlations with the MZR residuals (e.g., \FgastotE, and \Tefty, $\varrho\sim$0.2), introducing them as secondary relations does not reproduce significantly the scatter \citep[{ i.e., prove the need for} a secondary relation introduced by][]{Mannucci10}. This indicates { that parameters other than} the stellar mass may not been playing a significant role { in determining }the oxygen abundance.

\begin{figure*}[htp]
\centering
	\includegraphics[width=\textwidth]{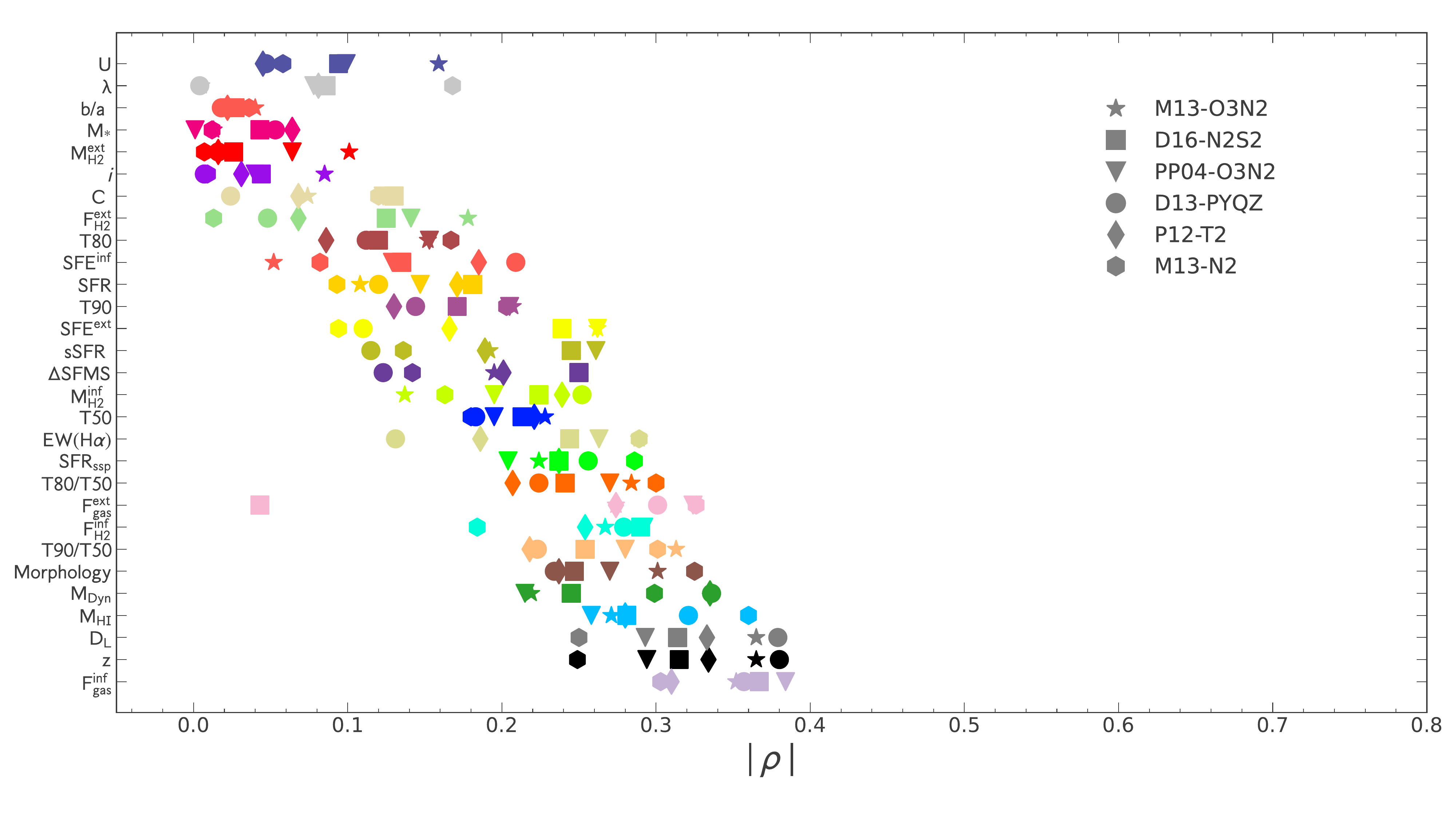}
\caption{Similar to Fig.~\ref{fig:cal_all}. The  symbols with no edge represent those correlation coefficients presented in Fig.~\ref{fig:slopes} derived between the residuals of the MZR, \DeltMZR, and the physical parameters of the galaxy other than the stellar mass. The significant reduction on the correlation coefficient for all the probed parameters with \DeltMZR\ suggests that the stellar mass is indeed the parameter that best describe the oxygen abundance.{We preserve the range of the correlation coefficient in absolute values to make the visual comparison between Figures \ref{fig:cal_all} and \ref{fig:cal_all_res} clearer.}
Furthermore, a comparison with Fig.~\ref{fig:cal_all} indicates that those large correlation factors derived for other physical parameters than the stellar mass using the absolute value of the oxygen abundance (edged symbols) are heavily induced by the stellar mass itself.}
\label{fig:cal_all_res}
\end{figure*}
As an additional { analysis, we} compare the strength of the correlation of the explored parameters with the oxygen abundance (Fig.~\ref{fig:cal_all}), and with the residual once removed the best fitted MZR relation (Fig.~\ref{fig:cal_all_res}). In contrast to the large correlation coefficients found for the oxygen abundance with the set of parameters explored in Sec.~\ref{sec:corr_coefs}, the coefficients of \DeltMZR\ (i.e., once we removed the dependence of \Mstar\ with the oxygen abundance) with these parameters are significantly smaller. Once again this comparison highlights the fact that \Mstar\ is the parameter that best describes the oxygen abundance in a galaxy. Furthermore, the stark difference between the two sets of coefficients is a strong suggestion { that, after removing the dependence} with the stellar mass, the other parameters do not appear to have a significant impact in shaping the chemical abundance in these galaxies. This difference also points out that those large correlation factors derived in Sec.~\ref{sec:corr_coefs} could be induced due to the significant correlation that those parameters have with the total stellar mass.

Despite the mild correlation coefficients observed between \DeltMZR\ and the different galaxy's properties \mbox{{ ($|\rho| \sim 0.4$)}}, there are still some parameters that may have a possible secondary impact on shaping the characteristic oxygen abundance in star-forming galaxies. Apart from the observed gas fraction, \FgastotI, the parameters that correlate the most with \DeltMZR\ are related to the projected distance of the galaxies (i.e., $z$, and \DL), suggesting a possible evolution of the characteristic oxygen abundance. We also note, that other parameters also related to the evolution of galaxy could also play a role in shaping this abundance (e.g., Morphology, \Tnfty, and \Tefty). 
On the contrary, we do not find significant correlations between \DeltMZR\ and those parameters usually associated with a strong reduction of the scatter of the MZR (e.g., SFR, and SFE). Finally, we note that although \FgastotI have a mild impact on setting the oxygen abundance, we require a larger an homogeneous set of \hi\ observations to truly asses its impact in the chemical enrichment of galaxies.

Another point to highlight is the slopes, and $\Delta{\sigma} / \sigma_{res}$ values yielded by the residuals for the observed and proxy data (i.e. \MmolExt(\MmolInf),\ \FmolExt(\FmolInf),\ \FgastotE(\FgastotI),\ \SFEExt(\SFEInf)), possibly are associated with mismatches between apertures, or with the method to determine proxy values, or even related to the size of the EDGE-APEX or HI datasets and the eCALIFA sample.

\subsection{The shape of the MZR}
\label{disc:shapeMZR}

As we mention above, in Sec.~\ref{sec:MZR} we derive the best shape of the MZR for our sample of star-forming galaxies. We find that the statistical treatment that provides the best shape of the MZR is when no bin is applied to the data, although the differences are subtle. The stellar mass binning forces the same statistical weight across the entire mass range. This may affect somehow the functional form of the MZR since the sample sizes are lower for both low and high mass regimes. Finally, for the un-binned data the OLS fitting provides the best results, while for the binned data applying an OLS or an ODR produce very similar results. Despite this, we find, regardless the metallicity calibrator, that the best functional form of the MZR is a third-order polynomial. This result contrasts with previously published studies indicating that the functional form of the MZR varies significantly depending to the calibrator \citep[e.g.,][]{Kewley2008}. Finally, we note that the stellar mass at which the highest oxygen abundance is reached depends on the calibrator (see Tab.~\ref{Tab:BestFit_paramCalib}).

Depending on the coefficients, a cubic polynomial could describe a function with two plateaus, at low and high values of the $x$-axis and a monotonous increase of $\mathrm{12 + log(O/H)}$ at intermediate stellar masses. 

According to the literature, those MZR features have possible physical explanations. The high mass saturation is linked to (1) effects of outflows at high mass \citep[][]{Tremonti04}; (2) a maximum yield \citep[see e.g.,][]{2007MNRAS.376..353P, Sanchez19}; (3) end of enrichment process at high mass, i.e., the effect of the SFHs \citep[e.g.,][]{vale-asari09, Camps21},
or (4) just an artifact from the fact that calibrators depending on [OIII] and [NII] are unable to trace higher abundances. Moreover, the increase of characteristic oxygen abundance at intermediate stellar masses, is related to consumption of the gas through the star formation process \citep[e.g.,][]{mcclure68, Lequeux79,Hughes13, Sanders15,Guo_2016}. Finally, the existence of a second plateau at the low mass regime could be caused due to the low efficiency of star formation \citep{Rossi06, Brooks_2007, Mouhcine08, Tassis08, Calura09}. 

As we show in Sec.~\ref{sec:MZR}, the best MZR fit show a plateau in the oxygen abundance for massive galaxies regardless the calibrator (see Fig.~\ref{fig:MZR_calibs}). However, depending on the calibrator we may find a mild flattening for the low-mass regime \citep[similar to the one reported using the SAMI or SDSS DR7 dataset,][]{Sanchez19,Blanc_2019}. Furthermore, we find an increasing trend at intermediate stellar masses regardless of the calibrator, that agrees with the general conception of the behavior of the oxygen abundance: more metal content in more massive galaxies. 

From the Monte Carlo analysis, we note that at the low-mass regime, the shape of the MZR is sensitive to variations with respect to the data uncertainty (see Fig.~\ref{fig:MZR_comparison}). We suggest that these differences could be caused due to the small statistics for this regime of \Mstar. A reliable estimation of the plateau of oxygen abundance at low-stellar masses has to be tested with a larger sample of galaxies in this mass regime of star-forming galaxies. Indeed, several studies of dwarf galaxies shows the opposite trend, with oxygen abundance still decreasing for masses below $\mathrm{log({M_{\star}/M{\odot}})}\sim 8$ \citep[e.g.,][]{2006ApJ...647..970L,2017A&A...601A..95C,Blanc_2019}. 
If this plateau is an actual feature of the MZR thus the chemical evolutionary models have to be able to explain the physical scenario that describes a constant oxygen abundances for both low and mass regimes.


\subsection{What drives the chemical enrichment in galaxies?}
\label{disc:chem_enri}

According to several chemical models, such as the one developed by \cite{Lilly13}, the chemical enrichment in a galaxy is mostly regulated by \Fmol\ and the current SFR. Broadly speaking, a large gas fraction suggests a metal-poor galaxy whereas metal-enriched galaxies are those with small \Fmol. As we note in Sec.~\ref{disc:corr}, the oxygen abundance shows strong correlations with different physical parameters, other than the stellar mass, including, \Fgastot, \Teighty, \Morph, SFR, among others. However, once we remove the dependence of the stellar mass with the metallicity we do not find those large correlation coefficients. In particular to those parameters usually associated to have a significant impact in the chemical enrichment of a star-forming galaxy such as \Fmol\ or SFR. Instead, we find that the residuals of the MZR correlate -- mildly -- with stellar properties (e.g., \Tfifty, and \Tninety).

Possible secondary relations between the oxygen abundance and properties different from the stellar mass could be explained since the oxygen abundance in the ISM is the byproduct of the stellar chemical evolution across the star formation history of a galaxy, SFH. In this regard, \Mstar\ can be thought as the final product of the SFH. Thus, the MZR and those subsequent correlations between \DeltMZR\ and the stellar properties could suggest that the SFH and the chemical history of a galaxy have a significant impact in shaping the oxygen abundance of the ISM instead of the current SFR. This agree with the recent results by \citet{Camps21}, where it is shown a clear evolution between the SFH and the Chemical enrichment histories in galaxies. 

Nevertheless, we cannot rule out the role of the cold component of the ISM in shaping the oxygen abundance. We find a mild correlation between \DeltMZR\ and \Fgastot\ (not with the molecular gas). Even more, there is also a correlation of \DeltMZR\ with morphology. Finally, previous studies using spatially resolved data suggest that the gas fraction plays a significant role in shaping the oxygen abundance at kpc scales  \citep[e.g.,][]{2014A&A...563A..58T,BB18}. Thus, the results of our analysis suggest a complex scenario for the chemical evolution in star-forming galaxies in the nearby universe. Although it is evident that the stellar mass is the main parameter that drives the oxygen abundance, there are different parameters that could potentially impact how galaxies modulate their chemical enrichment. From their SFH to the amount of cold gas in their ISM. This reinforces the idea that chemical models have to consider these parameters to provide a reliable description of the chemical abundance in galaxies.

\section{Conclusions}
\label{sec:Conclusions}

In this study we explore the correlation between a wide range of physical parameters and the characteristic oxygen abundance (measured at the effective radius) for a sample of 299 star-forming galaxies included in the eCALIFA survey. We adopted the most agnostic approach by performing an analysis using a heterogeneous set of six oxygen abundance calibrators. The physical parameters include properties from the stellar component, as well as, the cold and ionized components of the ISM. The main results of this study are as follows:
\begin{itemize}
    \item[$\diamond$] We confirm that the physical parameter that best correlates with the oxygen abundance is the stellar mass. Furthermore, we find that there are other physical parameters that could potentially have an impact in setting the oxygen abundance (see Fig.~\ref{fig:cal_all}).
    
    \item[$\diamond$]  We find that, regardless the calibrator, a third-order polynomial provides the best functional form that describes the Mass-Metallicity relation (MZR) for our dataset (see Figs.~\ref{fig:m13B1},~\ref{fig:m13B2}). 
    
    \item[$\diamond$] The residuals of the MZR present weaker correlations with any physical parameter than the oxygen abundance itself. Indeed, the correlations with this residual are weak even for those parameters related to the cold gas component (e.g., molecular gas fraction) and the SFR. On the other hand, those parameters related to the stellar component are those with the larger correlations with the MZR residuals (see Figs.~\ref{fig:m13B3},~\ref{fig:cal_all_res}).   
 
    \item[$\diamond$] Without prior judgment on which parameter produces a significant effect on the MZR and to what extent, we found that the parameters related to the SFH have equal or stronger importance than those usually explored. 
 
\end{itemize}

In summary, we suggest that, at least in star-forming galaxies, the chemical enrichment of the ISM is tightly connected to the star-formation history. Thus, those parameters more strongly connected with this SFH, such as the stellar mass, \Tfifty/\Teighty, or the morphology, are those that describe better the observed oxygen abundance. However, better estimations of the cold atomic gas are need to better constrain the effect of the total gas fraction in these processes.

\acknowledgments
We acknowledge support from the grants IA-100420 and IN100519 (DGAPA-PAPIIT ,UNAM) and funding from the CONACYT grants CF19-39578, CB-285080, and FC-2016-01-1916.

E.A.O acknowledge support from the SECTEI (Secretaría de Educación, Ciencia, Tecnología e Innovación de la Ciudad de México) under the Postdoctoral Fellowship SECTEI/170/2021 and CM-SECTEI/303/2021.

J.B-B also acknowledge support from the grant IA-101522 (DGAPA-PAPIIT ,UNAM).

\appendix

\section{Oxygen abundance calibrators}
\label{app_calib}

There is an extensive literature on the estimation of the oxygen abundance for galaxies, covering different methods, samples, and techniques. Along this study we have adopted a set of oxygen abundance calibrators of different nature { that include the largest number of galaxies}. They all use a set of line ratios (indexes) sensitive to the oxygen abundance defined as O2, O3, N2, S2, O3N2 and R23. They are defined as $ O2 \equiv log(\otwo\lambda3727/\hb)$, $ O3 \equiv log(\othree\lambda5007/\hb)$, $N2 \equiv log(\ntwo/\ha)$, $S2 \equiv log(\stwo/\ha)$, $ O3N2 \equiv log(\othree\lambda5007/\hb \times \ha/\ntwo\lambda6584)$,
and $R23\equiv (\otwo\lambda3727 +\othree\lambda4959+\othree\lambda5007)/ (\hb)$. Some calibrators use further combination of them, like O3O2($=$O3-O2) and N2S2($=$N2S2). We summarize here the main properties of these calibrators: 

\textbf{ M13-O3N2}. This calibrator is anchored to the “direct method” based on O3N2 index. To derive this calibration, \citet{marino13} comprises Te-based abundances of 603 HII regions extracted from the literature using a handful set of 16 Te-based HII regions provided by the CALIFA survey \citep{Sanchez13}. The oxygen abundance depends on the O3N2 line ratio in the following form is $12 + log(O/H) = 8.533[\pm0.012] - 0.214[\pm0.012] \cdot O3N2$.  It has an applicability interval of $-1.1<O3N2<1.7$, with a typical error of $\sim$0.09 dex.

\textbf{{ D16-N2S2}}. Calibrator based on a photoionization models for HII regions based on the $\ha$, [NII], and [SII] emission-lines \citep{dop16}. This metallicity calibration is claim to be independent of the ionization parameter and the reddening. It considers a 5th polynomial function in the form: $12+log(O/H)=8.77+y+0.45(y+0.3)^5$, with $y=log([NII]/[SII] + 0.264 log([NII]/\ha))$

\textbf{{ PP04-O3N2}}. Linear calibrator based on the O3N2 indicator derived by \citet{pp04} using a combination 137 extragalactic HII regions (for the low metallicity range) and a set of photoionization models (for the high metallicity range), valid within $−1 < O3N2 < 1.9$: $12 + log(O/H) = 8.73 - 0.32 \cdot O3N2$, with a typical error of $\sim$0.18 dex.

\textbf{{ D13-PYQZ}}. Oxygen abundance derived by the pyqz code \citep{PYQZ}, that employs pure photoionization models, using the O2, N2, S2, O3O2, O3N2, N2S2, and O3S2 line ratios.

\textbf{{ P12-T2}}. Calibrator based on a combination of different calibrators using measurements anchored to the direct method, such as M13-O3N2 and M13-N2, corrected by the effect of the  electron temperature inhomogeneity parameter \citep{T2}. The details of this calibrator are described in \citet{Sanchez19}.

\textbf{M13-N2}. Calibrator based on the N2 index derived by \citet{marino13}, using the same dataset described before for the M13-O3N2 calibrator. It adopts the linear form $12 + log(O/H) = 8.743[\pm0.027] - 0.462[\pm0.024] \cdot N2$, being valid for the interval $−1.6<N2<−0.2$, with a typical error of $\sim$0.09 dex.

In summary, we include strong-line calibrators anchored to the direct method (M13-O3N2 and M13-N2), corrected by the possible in-homogeneity of the temperature distribution (P12-T2), derived using photoionisation models (D16-N2S2 and D13-PYQZ), and finally hybrid calibrators, that use both measurements using the direct method and results by photoionization models (PP04-O3N2). Although we explore other abundance calibrators \citep[e.g.,][]{RyS} to construct the MZR, those calibrators yield a significant smaller number of galaxies in comparison to those calibrators presented here ($\sim$ 30\% less). Thus, this heterogeneous set of calibrators cover a wide range of flavors for the derivation of the oxygen abundance.

\section{Parameters for the different MZR fits based on M13-O3N2 calibrator.}
\label{app:24_fit_param_m13}
According to the functional forms and criteria mention in Sec.~\ref{sec:MZR}, we include the fitting parameters for each functional form explored for this work in Table~ \ref{FitParam_M13}.

\setcounter{table}{0}
\renewcommand{\thetable}{\thesection.\arabic{table}}
\begin{deluxetable*}{lcccccccccccc}[htp]
\tablecaption{Fit parameters for M13-O3N2 calibrator\label{FitParam_M13}}
\tablewidth{0pt}
\tabletypesize{\scriptsize}
\tablehead{
\colhead{Fit} & \colhead{Technique} & \colhead{Binning} & 
\colhead{Sample size} & \colhead{a} & 
\colhead{b} & \colhead{c} & 
\colhead{d} & \colhead{$\varrho_{\Delta{MZR}}$} & \colhead{BIC} & 
\colhead{$\tilde\chi^{2}_{MZR}$} & \colhead{$\sigma_{\Delta{MZR}}$}  & 
\colhead{$\Delta\sigma / \sigma$}  \\ 
\colhead{} & \colhead{} & \colhead{} & \colhead{(\%)} & \colhead{} & 
\colhead{} & \colhead{} & \colhead{} &
\colhead{} & \colhead{} & \colhead{} & \colhead{(dex)} & \colhead{(\%)} } 
\startdata
Linear & OLS & No & 100 & 7.24 $\pm$ 0.1 & 0.12 $\pm$ 0.1 & - & - & -0.13 & -773.9 & 0.517 & 0.064 & 28.9 \\
Linear & ODR & No & 100 & 6.42 $\pm$ 0.1 & 0.2 $\pm$ 0.1 & - & - & -0.6 & -664.7 & 0.752 & 0.077 & 14.4 \\
Linear & OLS & Yes & 100 & 6.88 $\pm$ 0.1 & 0.16 $\pm$ 0.1 & - & - & -0.37 & -748.0 & 0.565 & 0.067 & 25.6 \\
Linear & ODR & Yes & 100 & 7.06 $\pm$ 0.1 & 0.14 $\pm$ 0.1 & - & - & -0.26 & -765.8 & 0.532 & 0.065 & 27.8 \\
Linear & OLS & No & 90 & 7.24 $\pm$ 0.1 & 0.12 $\pm$ 0.1 & - & - & -0.13 & -773.9 & 0.517 & 0.064 & 28.9 \\
Linear & ODR & No & 90 & 6.42 $\pm$ 0.1 & 0.2 $\pm$ 0.1 & - & - & -0.6 & -664.7 & 0.752 & 0.077 & 14.4 \\
Linear & OLS & Yes & 90 & 6.88 $\pm$ 0.1 & 0.16 $\pm$ 0.1 & - & - & -0.37 & -748.0 & 0.565 & 0.067 & 25.6 \\
Linear & ODR & Yes & 90 & 7.06 $\pm$ 0.1 & 0.14 $\pm$ 0.1 & - & - & -0.26 & -765.8 & 0.532 & 0.065 & 27.8 \\ \hline
Exponential & OLS & No & 100 & 8.63 $\pm$ 0.1 & -17.57 $\pm$ 0.4 & - & - & -0.1 & -769.8 & 0.525 & 0.065 & 27.8 \\
Exponential & ODR & No & 100 & 8.72 $\pm$ 0.1 & -29.08 $\pm$ 0.5 & - & - & -0.54 & -581.1 & 0.994 & 0.089 & 1.1 \\
Exponential & OLS & Yes & 100 & 8.65 $\pm$ 0.1 & -19.08 $\pm$ 1.1 & - & - & -0.16 & -755.4 & 0.551 & 0.066 & 26.7 \\
Exponential & ODR & Yes & 100 & 8.65 $\pm$ 0.1 & -18.49 $\pm$ 1.0 & - & - & -0.14 & -761.6 & 0.541 & 0.066 & 26.7 \\
Exponential & OLS & No & 90 & 8.63 $\pm$ 0.1 & -16.36 $\pm$ 0.4 & - & - & -0.05 & -777.3 & 0.514 & 0.064 & 28.9 \\
Exponential & ODR & No & 90 & 8.72 $\pm$ 0.1 & -29.57 $\pm$ 0.6 & - & - & -0.56 & -571.1 & 1.03 & 0.09 & -0.0 \\
Exponential & OLS & Yes & 90 & 8.64 $\pm$ 0.1 & -17.97 $\pm$ 1.0 & - & - & -0.12 & -766.5 & 0.534 & 0.065 & 27.8 \\
Exponential & ODR & Yes & 90 & 8.63 $\pm$ 0.1 & -16.65 $\pm$ 1.1 & - & - & -0.06 & -775.8 & 0.519 & 0.064 & 28.9 \\ \hline
Polynomial & OLS & No & 100 & 8.72 $\pm$ 10.2 & -1.07 $\pm$ 3.0 & 0.19 $\pm$ 0.3 & -0.01 $\pm$ 0.1 & -0.04 & -789.4 & 0.495 & 0.063 & 30.0 \\
Polynomial & ODR & No & 100 & 231.45 $\pm$ 21.2 & -67.12 $\pm$ 6.3 & 6.7 $\pm$ 0.6 & -0.22 $\pm$ 0.1 & -0.28 & -534.6 & 1.169 & 0.096 & -6.7 \\
Polynomial & OLS & Yes & 100 & 55.15 $\pm$ 26.6 & -14.95 $\pm$ 7.9 & 1.57 $\pm$ 0.8 & -0.05 $\pm$ 0.1 & -0.11 & -790.1 & 0.5 & 0.062 & 31.1 \\
Polynomial & ODR & Yes & 100 & 8.01 $\pm$ 28.1 & -0.99 $\pm$ 8.2 & 0.19 $\pm$ 0.8 & -0.01 $\pm$ 0.1 & -0.1 & -773.2 & 0.525 & 0.064 & 28.9 \\
Polynomial & OLS & No & 90 & -12.32 $\pm$ 15.7 & 5.13 $\pm$ 4.5 & -0.42 $\pm$ 0.4 & 0.01 $\pm$ 0.1 & -0.04 & -765.7 & 0.536 & 0.065 & 27.8 \\
Polynomial & ODR & No & 90 & 300.96 $\pm$ 55.5 & -86.51 $\pm$ 16.1 & 8.5 $\pm$ 1.6 & -0.28 $\pm$ 0.1 & -0.19 & -345.0 & 2.241 & 0.133 & -47.8 \\
Polynomial & OLS & Yes & 90 & -37.01 $\pm$ 37.6 & 11.98 $\pm$ 10.8 & -1.05 $\pm$ 1.0 & 0.03 $\pm$ 0.1 & -0.1 & -668.1 & 0.746 & 0.077 & 14.4 \\
Polynomial & ODR & Yes & 90 & -10.05 $\pm$ 32.3 & 4.23 $\pm$ 9.3 & -0.31 $\pm$ 0.9 & 0.01 $\pm$ 0.1 & -0.1 & -739.2 & 0.59 & 0.068 & 24.4 \\
\enddata
\end{deluxetable*}

\section{Best MZR derivation for different calibrators}
\label{app:MZR_calibs}

\renewcommand\thefigure{\thesection.\arabic{figure}} 
\setcounter{figure}{0} 
\begin{figure*}[htp]
\centering
	\includegraphics[width=\textwidth]{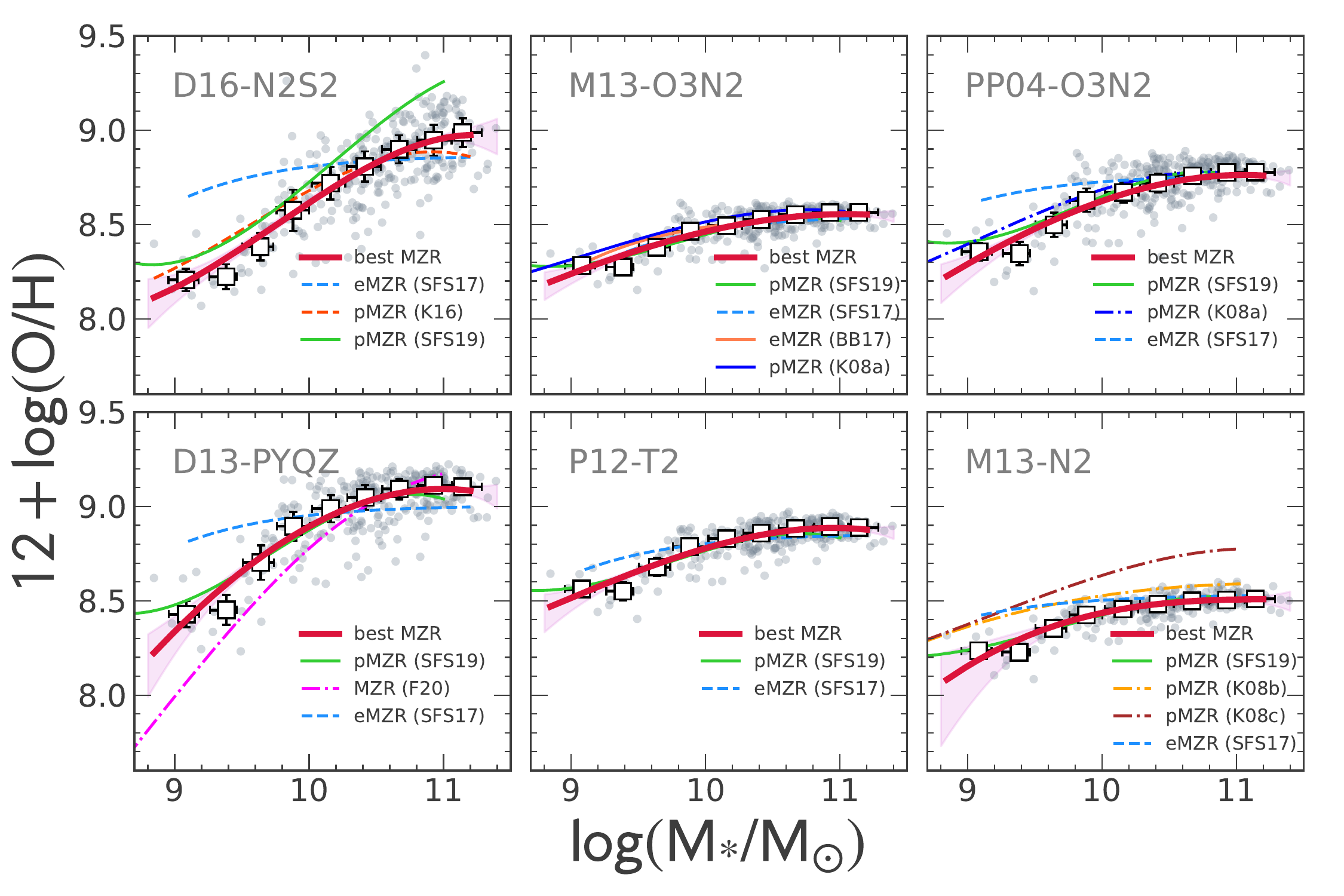}
	\caption{
	Distribution of oxygen abundances along the stellar mass for our sample of SFGs (grey circles) for the six different calibrators adopted along this study (each panel) together with the best fitted relation between them (3rd-order polynomial, red-solid line). The filled-white squares with error bars represent the average metallicity for stellar mass bin of 0.25 dex. For comparison purposes, we include, for each calibrator, exponential and polynomial MZR fits derived by literature studies based on single-fiber spectroscopy and IFS datasets, such as: (i) \citet{Kewley2008}, K08a, K08b, based on \ntwo\ index from \citealt{pp04} and K08c \citep{Kewley2008} based on \citealt{d02} calibrator, all them derived using the SDSS-DR4 data; (ii) \citet{2016ApJ...823L..24K}, K16, based on D16-N2S2 calibrator derived for SDSS-DR7 dataset; (iii) \citep{Sanchez17}, SFS17, and \citep{Sanchez19}, SFS19, based on different calibrators using the CALIFA and SAMI samples;  and finally (iv) \citep{BB17,2020ApJ...895..106F}, BB17 and F20, for the { M13-O3N2} and { D13-PYQZ} calibrators using MaNGA and MUSE samples respectively.}
\label{fig:MZR_calibs}
\end{figure*}

Previous studies using unresolved spectroscopic data have reported that the shape of the MZR heavily depends on the adopted calibrator \citep[See][]{Ellison08}. To study this in detail with resolved data, we repeat the analysis described in Sec.\ref{sec:MZR} using the M13-O3N2 calibrator with the heterogeneous set of oxygen abundance calibrators described before (Appendix \ref{app_calib}). 

From this analysis we conclude that regardless the calibrator, the stellar mass is the primary physical parameter that correlates with metallicity. 

Following the same procedure as in Sec.~\ref{sec:MZR}, we find that a 3rd-order polynomial using a unbinned scheme offers a reliable fit to the MZR, for 5 of the 6 tested calibrators (except for  M13-N2). 

Like in the case of the M13-O3N2 calibrator, the residuals for each calibrator using this functional form yield the lowest dependence with the stellar mass. Furthermore an OLS fitting technique provides the best fit for 5 from 6 calibrators (except for M13-N2). In summary there is no dependence of this result on the adopted calibrator.

\setcounter{table}{0}
\renewcommand{\thetable}{\thesection.\arabic{table}}
\begin{deluxetable*}{lcccccccccccc}[htp]
\tablecaption{Best fits parameter per calibrator}
\label{Tab:BestFit_paramCalib}
\tablewidth{0pt}
\tabletypesize{\scriptsize}
\tablehead{
\colhead{Calibrator} & \colhead{Fit} & 
\colhead{Sample size} & \colhead{A} & 
\colhead{B} & \colhead{C} & 
\colhead{D} & BIC & 
\colhead{$\tilde\chi^{2}_{MZR}$} & \colhead{$\sigma_{log(O/H)}$} & 
\colhead{$\sigma_{\Delta{MZR}}$} & \colhead{$\Delta\sigma / \sigma$} \\ 
\colhead{} & \colhead{} & \colhead{} & \colhead{} & 
\colhead{} & \colhead{} & \colhead{} &
\colhead{} & \colhead{} & \colhead{(dex)} & \colhead{(dex)} & \colhead{(\%)}
} 
\startdata
{ M13-O3N2} & OLS & 100 & 8.72 $\pm$ 10.2 & -1.07 $\pm$ 3.0 & 0.19 $\pm$ 0.3 & -0.01 $\pm$ 0.1 & -789.4 & 0.495 & 0.09 & 0.063 & 30.0 \\
{ D16-N2S2} & OLS & 100 & 71.89 $\pm$ 22.6 & -20.46 $\pm$ 6.7 & 2.15 $\pm$ 0.7 & -0.07 $\pm$ 0.1 & -242.9 & 0.406 & 0.249 & 0.157 & 37.0 \\
{ PP04-O3N2} & OLS & 100 & 8.95 $\pm$ 13.9 & -1.58 $\pm$ 4.1 & 0.28 $\pm$ 0.4 & -0.01 $\pm$ 0.1 & -551.3 & 0.494 & 0.142 & 0.094 & 34.0 \\
{ D13-PYQZ} & OLS & 100 & -1.23 $\pm$ 22.3 & 0.47 $\pm$ 6.5 & 0.17 $\pm$ 0.6 & -0.01 $\pm$ 0.1 & -356.9 & 0.452 & 0.195 & 0.13 & 33.0 \\
{ P12-T2} & OLS & 100 & 20.63 $\pm$ 11.1 & -4.75 $\pm$ 3.2 & 0.57 $\pm$ 0.3 & -0.02 $\pm$ 0.1 & -756.5 & 0.43 & 0.102 & 0.066 & 35.0 \\
{ M13-N2} & ODR binned & 100 & -26.79 $\pm$ 20.8 & 9.21 $\pm$ 6.0 & -0.8 $\pm$ 0.6 & 0.02 $\pm$ 0.1 & -825.9 & 0.523 & 0.082 & 0.059 & 28.0 \\
\enddata
\end{deluxetable*}

In Table \ref{Tab:BestFit_paramCalib} we list the parameters associated with the best fit for each calibrator including the uncertainties. In this Table, we also list $\Delta\sigma / \sigma$, this is the relative difference between the scatter of the residuals of the MZR and the scatter of the observed metallicity distribution for each calibrator. $\Delta\sigma / \sigma$ shows that once we include the correlation between the stellar mass with the metallicity we find a significant reduction in the scatter ($\sim$ 30\%), irrespectively of the calibrator. Hence, this result is in agreement with our previous analysis, namely that stellar mass is a clear proxy of the oxygen abundance.

\section{Dependence of $\Delta{MZR}$ with additional parameters for different calibrators }
\label{app:D_MZR_calibs}

\setcounter{table}{0}
\renewcommand{\thetable}{\thesection.\arabic{table}}

\begin{table*}[htp]
\centering
\caption{Results of the fitting of $\Delta{MZR}$ as a function of different physical parameters for a set of oxygen abundance calibrators}
\label{tab:statistical_parameters}
\scalebox{0.9}{
\rotatebox{90}{
\begin{tabular}{cccccccccccccccc}
\hline
\hline
\\
Cal. & \FgastotE/(\FgastotI) & \redshiftz & \DL & \MHI & \Mdyn & \Morph & \Tnfty& \FmolExt(\FmolInf) & \Tefty & \SFRssp & \EWre & \Tfifty &\MmolExt/(\MmolInf) & \SFMS & \sSFR  \\ \hline
\\
\multicolumn{16}{c}{{ Spearman} coefficient ($\varrho$) between the \DeltMZR\ and each observable}  \\\hline
D16-N2S2 & -0.274 (-0.352) & -0.365 & -0.365 & -0.271 & -0.219 & -0.301 & 0.313 & 0.178 (-0.267) & 0.284 & -0.224 & -0.289 & -0.228 & 0.101 (-0.137) & -0.195 & -0.192 \\
M13-O3N2 & -0.299 (-0.367) & -0.315 & -0.314 & -0.281 & -0.245 & -0.247 & 0.254 & 0.125 (-0.290) & 0.241 & -0.237 & -0.244 & -0.213 & 0.026 (-0.224) & -0.250 & -0.245 \\
PP04-O3N2 & -0.324 (-0.384) & -0.294 & -0.293 & -0.258 & -0.215 & -0.270 & 0.280 & 0.141 (-0.292) & 0.270 & -0.204 & -0.263 & -0.195 & 0.064 (-0.195) & -0.250 & -0.261 \\
D13-PYQZ & -0.301 (-0.357) & -0.380 & -0.379 & -0.321 & -0.336 & -0.234 & 0.223 & 0.048 (-0.279) & 0.224 & -0.256 & -0.131 & -0.183 & -0.016 (-0.252) & -0.123 & -0.115 \\
P12-T2 & -0.274 (-0.310) & -0.334 & -0.333 & -0.280 & -0.335 & -0.237 & 0.218 & 0.068 (-0.254) & 0.207 & -0.237 & -0.186 & -0.221 & -0.016 (-0.239) & -0.201 & -0.189 \\
M13-N2 & -0.326 (-0.303) & -0.249 & -0.250 & -0.360 & -0.299 & -0.325 & 0.301 & 0.013 (-0.184) & 0.300 & -0.286 & -0.289 & -0.180 & 0.007 (-0.163) & -0.142 & -0.136\\\hline
Mean & -0.300 (-0.345) & -0.323 & -0.322 & -0.295 & -0.275 & -0.269 & 0.265 & 0.095 (-0.261) & 0.254 & -0.241 & -0.234 & -0.203 & 0.028 (-0.202) & -0.194 & -0.190 \\ \hline
\\
\multicolumn{16}{c}{Slopes ($m$) per calibrator calculated using the \DeltMZR\ against each parameter linear fit} \\\hline
D16-N2S2 & -0.216 (-0.145) & -7.200 & -0.295 & -0.110 & -0.063 & -0.021 & 0.350 & 0.079 (-0.191) & 0.250 & -0.097 & -0.180 & -0.162 & 0.033 (-0.073) & -0.058 & -0.107 \\
M13-O3N2 & -0.075 (-0.052) & -2.686 & -0.090 & -0.018 & -0.028 & -0.011 & 0.106 & 0.037 (-0.060) & 0.095 & -0.024 & -0.018 & -0.103 & -0.008 (-0.014) & -0.033 & -0.068 \\
PP04-O3N2 & -0.134 (-0.095) & -3.524 & -0.135 & -0.025 & -0.036 & -0.018 & 0.174 & 0.069 (-0.095) & 0.161 & -0.031 & -0.035 & -0.139 & -0.012 (-0.019) & -0.058 & -0.113 \\
D13-PYQZ & -0.150 (-0.098) & -6.441 & -0.241 & -0.063 & -0.067 & -0.018 & 0.184 & 0.021 (-0.143) & 0.183 & -0.074 & -0.034 & -0.273 & -0.032 (-0.047) & -0.009 & -0.082 \\
P12-T2 & -0.077 (-0.042) & -2.658 & -0.109 & -0.023 & -0.039 & -0.011 & 0.088 & 0.028 (-0.058) & 0.086 & -0.041 & 0.008 & -0.113 & -0.024 (-0.021) & -0.022 & -0.068 \\
M13-N2 & -0.076 (-0.026) & -1.514 & -0.072 & -0.019 & -0.035 & -0.005 & 0.069 & 0.003 (-0.054) & 0.061 & -0.026 & -0.030 & -0.168 & -0.015 (-0.022) & -0.011 & -0.020 \\\hline
Mean & -0.121 (-0.076) & -4.004 & -0.157 & -0.043 & -0.045 & -0.014 & 0.162 & 0.040 (-0.100) & 0.139 & -0.049 & -0.048 & -0.160 & -0.010 (-0.033) & -0.032 & -0.076 \\ \hline
\\
\multicolumn{16}{c}{Scatter reduction (${\Delta{\sigma}/ \sigma}_{res}$) per calibrator for each residual observable ($\%$)}\\\hline
D16-N2S2 & 0.9 (3.4) & 2.4 & 4.9 & 4.0 & 2.6 & 3.4 & 0.2 & 2.9 (1.8) & 3.6 & 2.1 & 4.2 & 1.5 & 0.5 (-0.1) & 1.6 & 1.5 \\
M13-O3N2 & 3.3 (5.9) & 0.2 & 4.2 & 3.0 & 4.1 & 1.1 & 1.5 & 2.9 (1.1) & 2.4 & 2.4 & 1.8 & 3.9 & -1.2 (0.5) & 2.7 & 1.2 \\
PP04-O3N2 & 4.6 (7.2) & 1.0 & 3.3 & 2.2 & 3.1 & 2.2 & 2.2 & 3.1 (1.0) & 3.3 & 1.6 & 2.3 & 2.9 & -1.5 (-0.1) & 2.9 & 1.7 \\
D13-PYQZ & 3.0 (6.7) & 1.9 & 5.6 & 5.0 & 5.1 & 0.6 & 1.4 & 0.8 (1.7) & 1.4 & 3.5 & 0.7 & 6.5 & -1.6 (-0.3) & 0.2 & -1.0 \\
P12-T2 & 4.1 (5.3) & 2.2 & 4.0 & 3.5 & 5.0 & 1.6 & 2.4 & 1.6 (1.0) & 2.3 & 2.3 & -0.7 & 4.2 & -4.5 (-0.7) & 1.5 & 0.1 \\
M13-N2 & 2.4 (4.9) & 1.5 & 2.6 & 4.5 & 2.5 & 1.2 & 3.0 & 0.3 (-1.5) & 2.2 & 2.8 & 2.4 & 12.3 & -1.6 (-4.3) & 0.1 & -0.6 \\ \hline
Mean & 3.1 (5.6) & 1.5 & 4.1 & 3.7 & 3.7 & 1.7 & 1.8 & 1.9 (0.8) & 2.5 & 2.4 & 1.8 & 5.2 & -1.7 (-0.8) & 1.5 & 0.5 \\ \hline
\end{tabular}}}
\end{table*}

Following the same analysis shown in  Sec. \ref{sec:Residuals}, we explore the possible correlation between the residuals of the MZR and those observables that show a significant correlation with metallicity ($\rho > 0.3$, see Fig.~\ref{fig:cal_all}). For each calibrator we compute the correlation coefficient between the residuals of the MZR ($\Delta{MZR}$) and each physical parameter ($\varrho$), the slope of a linear fit between $\Delta{MZR}$ and the physical parameter ($m$), and the relative change in the residuals once applied this linear relation ($\Delta \sigma / \sigma_{res}$). As indicated above, Fig. ~\ref{fig:slopes} summarizes the results of this analysis, which the individual parameters listed in Table \ref{tab:statistical_parameters}. As discussed above (see Fig. ~\ref{fig:cal_all_res} in Sec. \ref{sec:Residuals}), the correlation coefficients derived from the residuals of the MZR, suggests that no physical parameter presents a clear secondary correlation with the oxygen abundance, regardless of the calibrator.

\begin{figure*}[htp]
\centering
	\includegraphics[width=\textwidth]{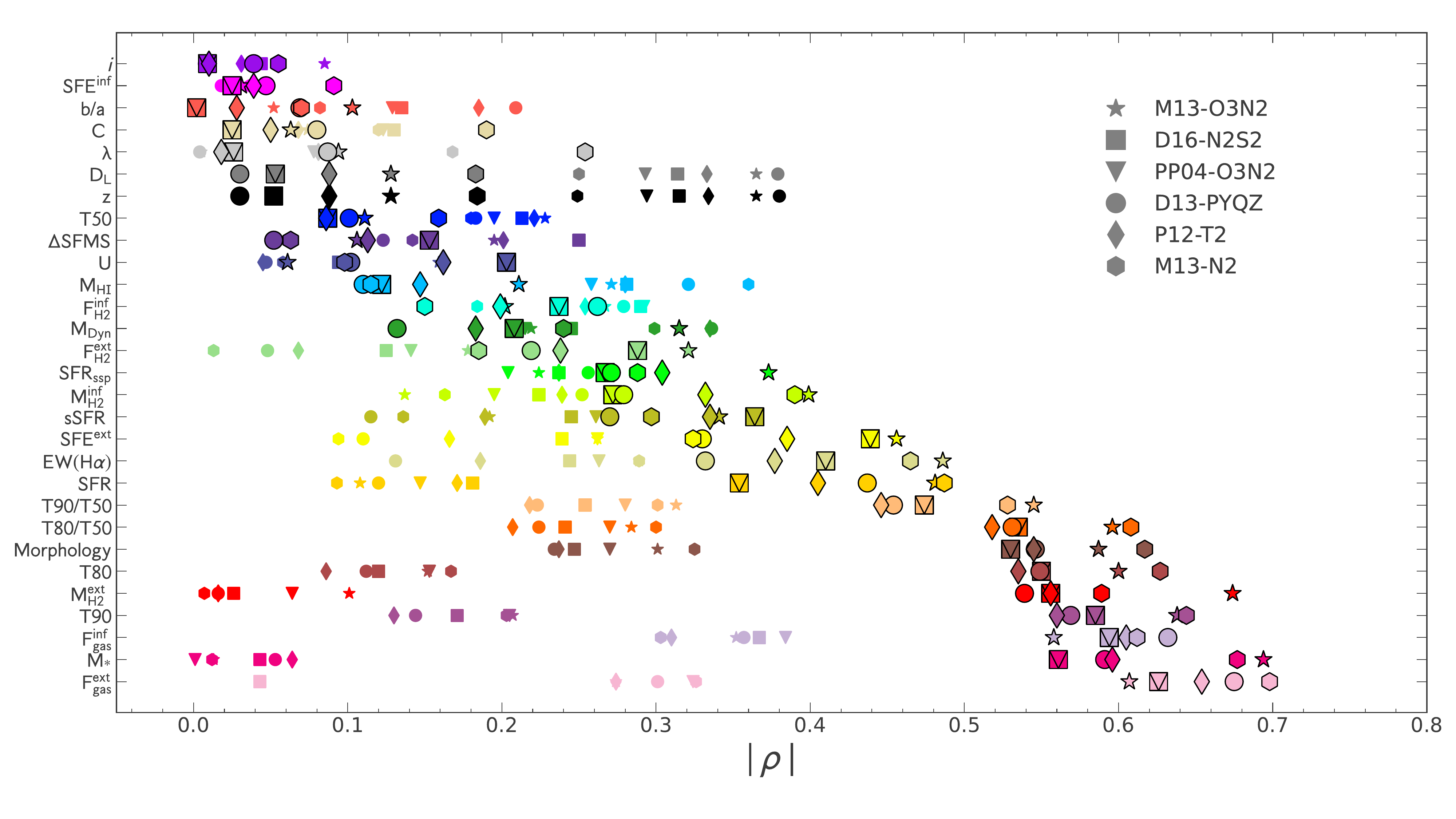}
\caption{ Correlation ranking for oxygen abundance with 29 parameters of the galaxies and residuals. We include the six calibrators used along this paper. The symbols with(out) black edges represent the correlations coefficient calculated with the parameters of galaxies(residuals).}
\label{fig:cal_coef_comparison}
\end{figure*}

In Fig.~\ref{fig:cal_coef_comparison} we use the same order of parameters presented in Fig.~\ref{fig:cal_all}, to rank the correlation coefficients calculated for residuals in Fig.~ \ref{fig:cal_coef_comparison}. This provides a visual comparison between each other. This figure highlights the fact that once removed the primary correlation of the oxygen abundance with \Mstar, there is no other parameter that exhibits a significant correlation with the residuals. Furthermore, those parameters that has a possible correlation with the residuals of the MZR are those that show weak correlation with the oxygen abundance itself (e.g., \FgastotI, $z$, and \DL).

\bibliography{sample63}{}
\bibliographystyle{aasjournal}
\end{document}